%% file: article.tex
\documentclass[review]{elsarticle}

\usepackage{setspace}
\usepackage{amssymb}
\usepackage{amsmath,graphicx,hyperref}
\usepackage{booktabs}
\usepackage{url}
\usepackage{tcolorbox}
\usepackage{tabularx}
\usepackage{multirow}
\usepackage{subcaption}
\usepackage{lineno}
\usepackage{etoolbox}

\AtBeginEnvironment{equation}{%
\setlength{\abovedisplayskip}{2pt}%
\setlength{\belowdisplayskip}{2pt}%
}
\AtBeginEnvironment{align}{%
\setlength{\abovedisplayskip}{2pt}%
\setlength{\belowdisplayskip}{2pt}%
}

\journal{Signal Processing}

\begin{document}

\begin{frontmatter}

\title{Coupled Meta-Adaptive Filtering for Active Noise Control Under Time-Varying Acoustic Paths}

\author[ntu]{Boxiang Wang}
\author[ntu]{Zhengding Luo}
\author[ntu]{Ziyi Yang}
\author[npu]{Dongyuan Shi}
\author[zju]{Xuexian Liu}
\author[ntu]{Woon-Seng Gan}

\affiliation[ntu]{organization={Smart Nation TRANS Lab},
            addressline={School of Electrical and Electronic Engineering, Nanyang Technological University},
            city={Singapore},
            postcode={639798},
            country={Singapore}}

\affiliation[npu]{organization={Center of Intelligent Acoustics and Immersive Communications},
            addressline={School of Artificial Intelligence, Northwestern Polytechnical University},
            city={Xi'an},
            postcode={710072},
            country={China}}

\affiliation[zju]{organization={College of Energy Engineering},
            addressline={Zhejiang University},
            city={Hangzhou},
            postcode={310027},
            country={China}}

\begin{abstract}
Meta-adaptive filtering (Meta-AF) provides a data-driven alternative to hand-crafted adaptive filter updates by employing a learned optimizer throughout online adaptation. However, when Meta-AF is used for active noise control (ANC), time-varying acoustic paths remain a major challenge. In particular, the physics-informed optimizer features are constructed using a secondary path estimate and become mismatched when the physical path changes, leading to inaccurate filter updates and degraded noise reduction. To address this problem, this paper proposes a Coupled Meta-Adaptive Filtering Active Noise Control (CoMeta-AF-ANC) method, which applies meta-learning to jointly learn the control filter adaptation and acoustic path tracking within a closed-loop framework. The proposed Meta-Gated Joint Path Identifier (MG-JPI) simultaneously tracks the primary and secondary paths from the available ANC signals without auxiliary noise, while the updated secondary path estimate is fed back to reconstruct the Meta-AF controller features. A delayless dual-rate realization performs learned adaptation at the frame rate while generating the control signal at the sampling rate in the time domain. Evaluation using measured headrest acoustic paths shows that CoMeta-AF-ANC outperforms representative ANC algorithms in tracking time-varying acoustic paths, while maintaining higher stability across unseen head movement scenarios. It also generalizes well to real-world noises not encountered during training. \fntext[]{The code will be available at \href{https://github.com/Wang-Boxiang/CoMeta-AF-ANC}{https://github.com/Wang-Boxiang/CoMeta-AF-ANC}}
\end{abstract}

\begin{highlights}
\item A coupled meta-adaptive filtering framework is proposed for ANC under time-varying acoustic paths.
 \item A meta-gated joint path identifier enables auxiliary-noise-free joint primary and secondary path tracking.
\item  A delayless dual-rate realization decouples sample-rate control from frame-rate learned adaptation.
\item Experiments on measured acoustic paths demonstrate fast, stable tracking and generalization across unseen path variations and noises.
\end{highlights}


\end{frontmatter}

\input{sections/01_introduction.tex}
\input{sections/02_preliminaries.tex}

\input{sections/03_proposed_method.tex}
\input{sections/04_experiments.tex}

\input{sections/05_conclusion.tex}

\bibliographystyle{elsarticle-num}
\bibliography{refs}

\end{document}

%% file: sections/01_introduction.tex
\section{Introduction}
Active noise control (ANC) attenuates low-frequency noise by generating anti-noise from a secondary source to destructively interfere with the unwanted disturbance~\cite{elliott2000signal,kuo1999active,kajikawa2012recent}. Compared with passive methods that rely on bulky barriers, ANC provides a compact and effective solution and has been widely adopted in many applications~\cite{cheer2019application,chang2022multi,xiao2025soft}. Conventional ANC systems often employ LMS-based adaptive algorithms, such as the filtered-reference LMS (FxLMS) algorithm~\cite{yang2020stochastic}. However, algorithms based on such hand-crafted update rules inherently suffer from slow convergence, a risk of divergence, and limited tracking capability under rapidly changing acoustic conditions, while also requiring empirical parameter tuning~\cite{zhang2018active,iotov2025speech,tang2023stability}.

In recent years, deep learning has been introduced into ANC to overcome the limitations of conventional adaptive algorithms. Selective fixed-filter ANC (SFANC) selects the most suitable pre-trained control filter based on the noise characteristics~\cite{shi2022selective}, whereas generative fixed-filter ANC (GFANC) generates a more suitable control filter to further enhance noise reduction performance~\cite{luo2025mssp,
wang2026spatial}. Alternatively, deep neural networks (DNNs) have been employed as controllers to directly generate the control signal~\cite{zhang2021deep,cha2023dnoisenet}. Deep learning has also been combined with adaptive filtering to exploit the complementary strengths of offline learning and online adaptation. In such systems, adaptive filtering can refine the initial response provided by the DNN under distribution shifts~\cite{luo2022hybrid,aboutiman2025subjective,bai2026adaptive}, while DNNs can augment adaptive ANC with capabilities such as nonlinear acoustic path modeling~\cite{park2023had} and sound-field interpolation~\cite{zhang2024active}. Another line of work applies model-agnostic meta-learning (MAML) to accelerate the convergence of adaptive algorithms by learning the initialization of the control filter~\cite{shi2021fast,shen2025data}, the joint initialization of the control filter and secondary path~\cite{yang2026co}, and the FxLMS step size~\cite{li2026practical}. Nevertheless, most existing hybrid deep learning–adaptive approaches retain a modular design in which the DNN and adaptive filtering perform distinct roles, while the adaptive filter update itself remains governed by a conventional algorithm. 

A more direct strategy is to use a DNN as the online optimizer for adaptive filtering. Meta-adaptive filtering (Meta-AF) formulates adaptive filter design as a meta-learning problem, in which a DNN maps signal features directly to filter increments~\cite{casebeer2021auto}. By learning the filter update mechanism directly from data, Meta-AF reduces reliance on hand-crafted update rules and empirical parameter tuning, while achieving faster convergence than conventional algorithms across several audio adaptive-filtering tasks~\cite{casebeer2022meta}. Unlike MAML-based methods that use meta-learning to learn the initialization and subsequently employ conventional adaptive algorithms, Meta-AF keeps the learned optimizer active throughout online adaptation. This idea has been introduced into ANC to address nonlinear distortions, where the learned optimizer predicts an approximate gradient that is scaled by a prescribed step size to obtain the control filter update~\cite{feng2025meta}.

In practical ANC systems, the acoustic paths are often time-varying because of device repositioning~\cite{liebich2018signal}, listener motion~\cite{oh2026head}, or environmental changes~\cite{lian2023frequency}. Secondary path variation is especially critical for Meta-AF-based ANC because several physics-informed optimizer features are constructed using the secondary path estimate. When the secondary path changes, a mismatched model can therefore distort the features supplied to the learned optimizer and lead to inaccurate control filter updates, degrading both noise reduction and tracking performance. The existing Meta-AF-based ANC method relies on a fixed secondary path model~\cite{feng2025meta}. Although some robustness to moderate path mismatch has been reported, the method lacks online path tracking capability in time-varying acoustic environments.

Conventional online secondary path modeling (OSPM) provides a potential solution to this problem~\cite{wen2026asynchronous,ma2026robust,wang2025online}, but existing approaches introduce their own limitations. Auxiliary-noise-based OSPM methods inject an additional probing signal to maintain identification excitation, at the cost of introducing extra residual noise~\cite{eriksson1989use,zhang2003robust}. Auxiliary-noise-free OSPM methods instead exploit the controller output for secondary path modeling, but the identifiability of the secondary path depends on the spectral richness of the control signal~\cite{pradhan20205,hu2019online}. Sensor-conditioned neural approaches can infer path variations from contextual information but require additional devices such as cameras~\cite{oh2026head}, increasing system complexity and potentially raising privacy concerns.

Motivated by the challenges posed by time-varying acoustic paths, this paper proposes \emph{Coupled Meta-Adaptive Filtering Active Noise Control} (CoMeta-AF-ANC), which integrates control adaptation and online acoustic-path tracking into a unified meta-learning framework. The framework couples a meta-learned controller optimizer with a \emph{Meta-Gated Joint Path Identifier} (MG-JPI), which jointly tracks the primary and secondary paths using only the available ANC signals. The two adaptation processes are mutually dependent: the controller output provides the excitation required for path identification, while the updated secondary path estimate is immediately used to reconstruct the physics-informed features for subsequent controller adaptation. The controller and path optimizers are jointly trained through the same closed-loop recursion. In addition, a delayless dual-rate realization generates the control signal sample by sample in the time domain while performing learned adaptation at the frame rate, thereby preserving causal feedforward control without block-processing delay. The main contributions are summarized as follows:
\begin{itemize}
    \item We propose CoMeta-AF-ANC, a coupled meta-adaptive filtering framework for ANC under time-varying acoustic paths, in which learned optimizers jointly govern the online updates of the control filter and the primary and secondary path models within a causal closed-loop recursion.

     \item We introduce MG-JPI, an auxiliary-noise-free joint path identifier that tracks the primary and secondary paths using only the available ANC signals, enabling effective Meta-AF control during path variations.

    \item We realize the proposed framework using a delayless dual-rate structure, where the control signal is generated sample by sample while the learned control and path adaptation processes operate at the frame rate, thereby avoiding block processing delay in the control signal path.

    \item Extensive experiments using measured acoustic paths and diverse noise signals validate the effectiveness and generalization capability of the proposed CoMeta-AF-ANC against representative ANC baselines under time-varying acoustic conditions.
\end{itemize}

The remainder of this paper is organized as follows. Section~\ref{sec:preliminaries} introduces the feedforward ANC formulation and the direct extension of Meta-AF to ANC. Section~\ref{sec:proposed_cometaaf} presents the proposed CoMeta-AF-ANC framework, followed by the experimental evaluation in Section~\ref{sec:experiments}. Finally, Section~\ref{sec:conclusion} concludes the paper.

%% file: sections/02_preliminaries.tex
\section{Preliminaries}
\vspace*{-0.2cm}
\label{sec:preliminaries}
This section reviews the conventional FxLMS-based feedforward ANC formulation and the direct extension of Meta-AF to ANC. Their limitations are also discussed to motivate the proposed framework.

\vspace*{-0.2cm}
\subsection{FxLMS-Based Control Filter Adaptation}
\vspace*{-0.1cm}
\begin{figure}[!t]
\centering
\includegraphics[width=\columnwidth]{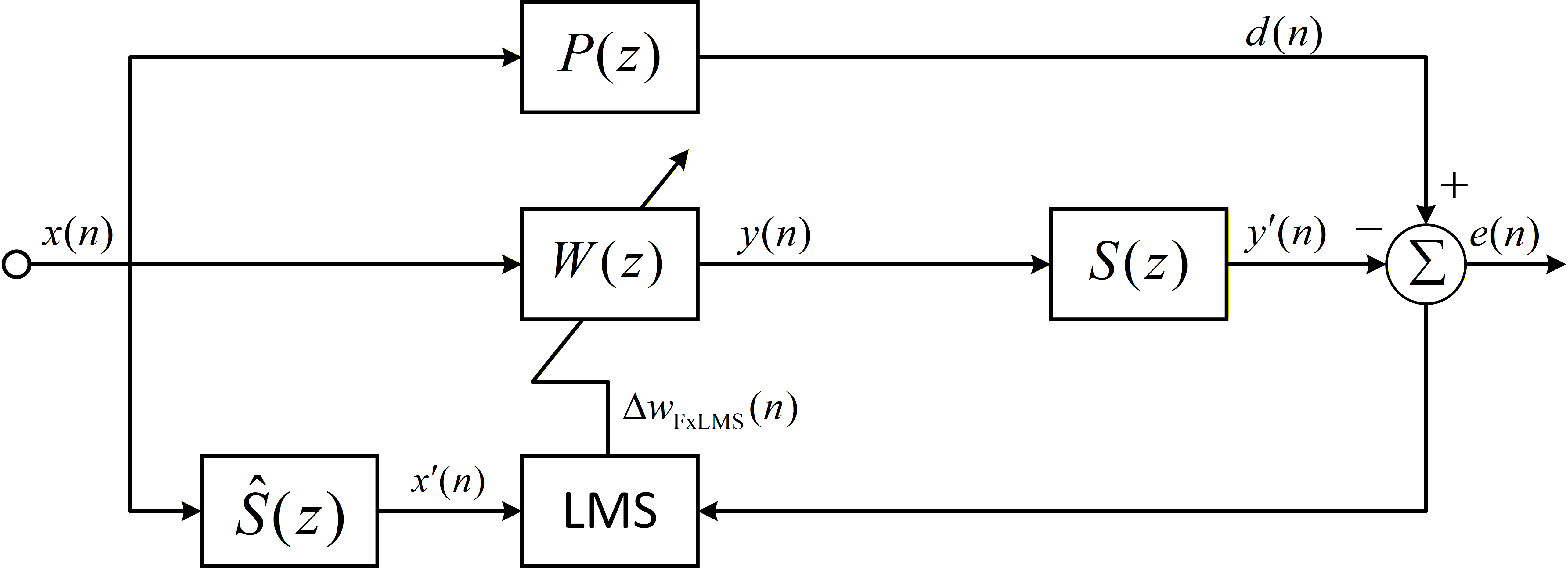}\vspace*{-0.2cm}
\caption{Block diagram of feedforward ANC with FxLMS-based control filter adaptation.}
\label{fig_1}
\end{figure}
Fig.~\ref{fig_1} illustrates a single-channel feedforward ANC system with the conventional FxLMS adaptation. Let $n$ denote the sample index. The control filter is represented by $\mathbf w(n)=[w_1(n),w_2(n),\ldots,w_{L_w}(n)]^{\mathrm T}$, and the corresponding reference signal vector is $\mathbf x(n)=[x(n),x(n-1),\ldots,x(n-L_w+1)]^{\mathrm T}$. Here, $L_w$ is the control filter length and $(\cdot)^{\mathrm T}$ denotes the transpose. The control signal driving the secondary source is
\begin{equation}
    y(n)=\mathbf w^{\mathrm T}(n)\mathbf x(n).
    \label{eq:loudspeaker_signal}
\end{equation}

The resulting error signal is
\begin{equation}
    e(n)=d(n)-y'(n)=p(n)*x(n)-s(n)*y(n),
    \label{eq:measured_error}
\end{equation}
where $p(n)$ and $s(n)$ denote the impulse responses of the primary and secondary paths, respectively, $d(n)$ is the disturbance at the error microphone, $y'(n)$ is the anti-noise reaching the error microphone, and $*$ denotes the convolution.

According to the FxLMS algorithm, the control filter vector is updated by
\begin{equation}
    \mathbf w(n+1)=\mathbf w(n)+\Delta\mathbf w_{\rm FxLMS}(n),
    \label{eq:fxlms_update}
\end{equation}
where the filter increment is
\begin{equation}
    \Delta\mathbf w_{\rm FxLMS}(n)=\mu e(n)\mathbf x'(n),
    \label{eq:fxlms_increment}
\end{equation}
with $\mu$ denoting the step size, and $\mathbf x'(n)=\hat s(n)*\mathbf x(n)$ is the filtered reference signal vector, where $\hat s(n)$ denotes the estimated secondary path impulse response.

However, FxLMS-based algorithms employ a hand-crafted update rule to compute the filter increment, suffer from slow convergence and poor tracking capability, and may diverge if the step size is not properly chosen. To alleviate these problems, the proposed CoMeta-AF-ANC method uses Meta-AF to update the control filter with the update rule directly learned from data.

\vspace*{-0.2cm}
\subsection{Meta-AF-Based Control Filter Adaptation}
\label{sec:metaaf_anc}
\begin{figure}[!t]
\centering
\includegraphics[width=\columnwidth]{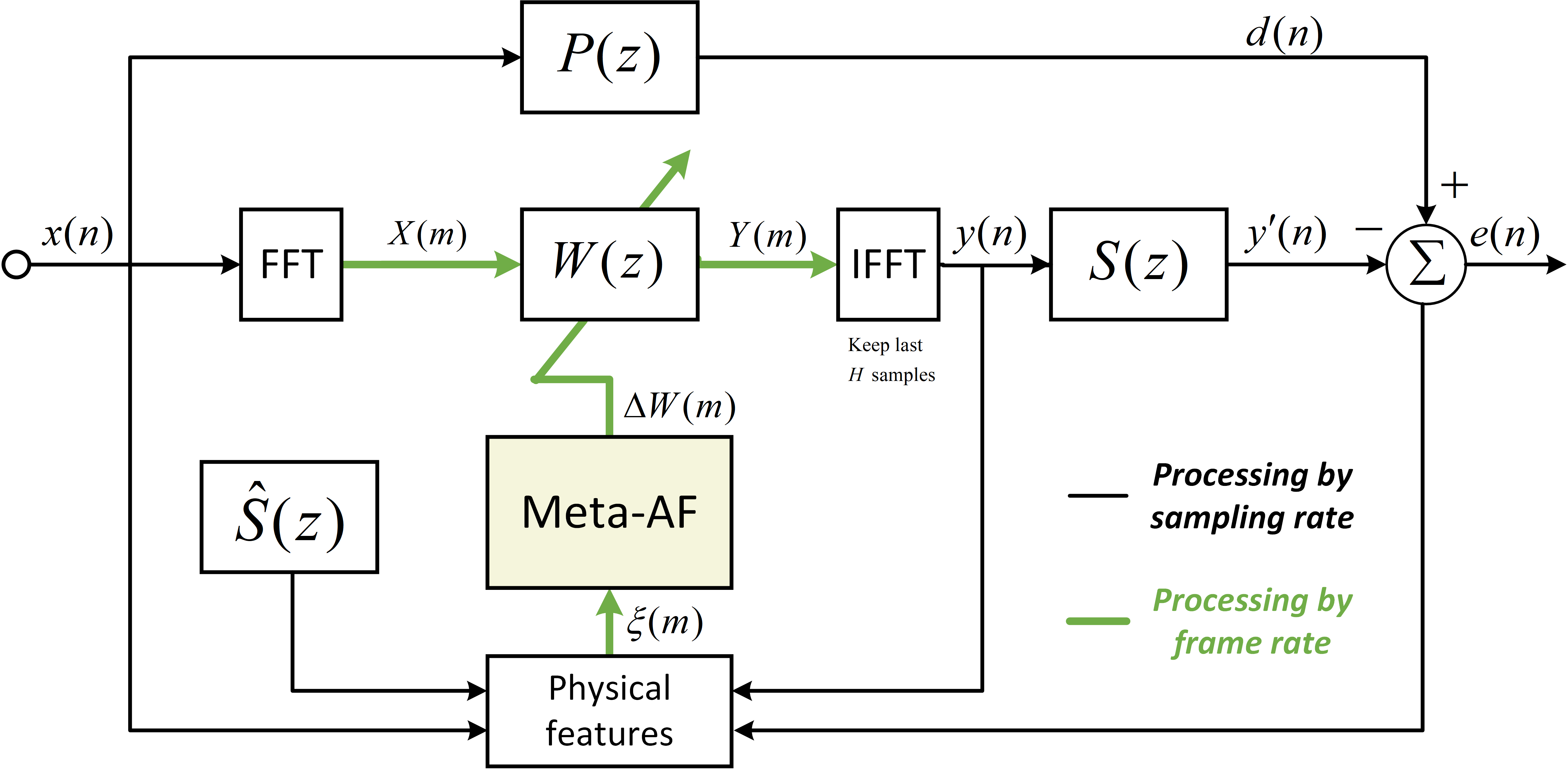}\vspace*{-0.2cm}
\caption{Block diagram of feedforward ANC with Meta-AF-based control filter adaptation.}
\label{fig_2}
\vspace*{-0.2cm}
\end{figure}

We first extend the Meta-AF optimizer to a feedforward ANC formulation that adopts a fixed secondary path estimate and generates the control signal in the frequency domain, as shown in Fig.~\ref{fig_2}. Unlike the Meta-AF-based ANC method in~\cite{feng2025meta}, which predicts an approximate control filter gradient and then applies a prescribed step size, this extension follows the original Meta-AF framework~\cite{casebeer2022meta} and directly predicts the control filter increment.

Let $m$ denote the frame index, $N$ the FFT/IFFT size, and $H$ the hop size. The control signal is generated in the frequency domain as
\begin{equation}
    Y(m)=W(m)X(m),
    \label{eq:metaaf_control_signal}
\end{equation}
where $X(m)$, $W(m)$, and $Y(m)$ denote the frequency-domain reference, control filter, and control signal, respectively. 

The control signal is transformed to the time domain by the inverse fast Fourier transform (IFFT), and the error signal is given by
\begin{equation}
    e(n)=d(n)-y'(n)=p(n)*x(n)-s(n)*\operatorname{IFFT}\!\left\{Y(m)\right\}.
    \label{eq:metaaf_measured_error}
\end{equation}
Following overlap-save processing, only the $H$ valid samples of each inverse-transformed block are retained.

The control filter is updated in the frequency domain as
\begin{equation}
    W(m+1)=W(m)+\Delta W_{\rm Meta}(m),
    \label{eq:generic_meta_af_update}
\end{equation}
where the control filter increment is produced by the Meta-AF optimizer as
\begin{equation}
    \Delta W_{\rm Meta}(m)=g_{\phi}\!\left(\boldsymbol\xi(m)\right).
    \label{eq:generic_meta_af}
\end{equation}
Here, $g_{\phi}(\cdot)$ denotes the Meta-AF optimizer parameterized by $\phi$. 

Following the original Meta-AF input design~\cite{casebeer2022meta}, the five features retain the same physical roles, namely the filter gradient, filter input, desired signal, filter output, and error. For feedforward ANC, the corresponding feature vector is
\begin{equation}
    \boldsymbol\xi(m)=\left[-X'^*(m)E(m),\;X'(m),\;\hat D(m),\;\hat Y'(m),\;E(m)\right]^{\mathrm T},
    \label{eq:metaaf_anc_features}
\end{equation}
where $(\cdot)^*$ denotes complex conjugation, $E(m)$ denotes the error signal spectrum, $X'(m)$ the filtered reference signal spectrum, $\hat D(m)$ the estimated disturbance spectrum, and $\hat Y'(m)$ the estimated anti-noise spectrum. For compactness, the frequency bin index is omitted.

Unlike FxLMS-based algorithms, which compute the filter increment using a hand-crafted update rule, Meta-AF predicts the increment directly from signal features. However, several of these features depend on the secondary path estimate. If the physical secondary path varies while the estimate remains fixed, the features supplied to the optimizer become mismatched, leading to inaccurate control filter updates. Moreover, block-wise frequency-domain control signal generation introduces additional processing delay. These limitations motivate the delayless realization and online secondary path tracking of the proposed CoMeta-AF-ANC method.

%% file: sections/03_proposed_method.tex
\vspace*{-0.3cm}
\section{The Proposed CoMeta-AF-ANC Method}
\vspace*{-0.2cm}
\label{sec:proposed_cometaaf}
This section details the proposed CoMeta-AF-ANC method. Unlike the formulation in Section~\ref{sec:metaaf_anc}, which generates the control signal in blocks via frequency-domain processing, CoMeta-AF-ANC generates the control signal sample by sample in the time domain while performing Meta-AF adaptation at the frame rate. We further introduce MG-JPI to jointly track the primary and secondary paths online, enabling the controller features to be updated according to the current acoustic paths. The controller and path adaptation processes are coupled through the ANC loop, where the controller output excites path identification and the updated secondary path estimate supports subsequent controller adaptation.

\begin{figure}[!t]
\centering
\includegraphics[width=\columnwidth]{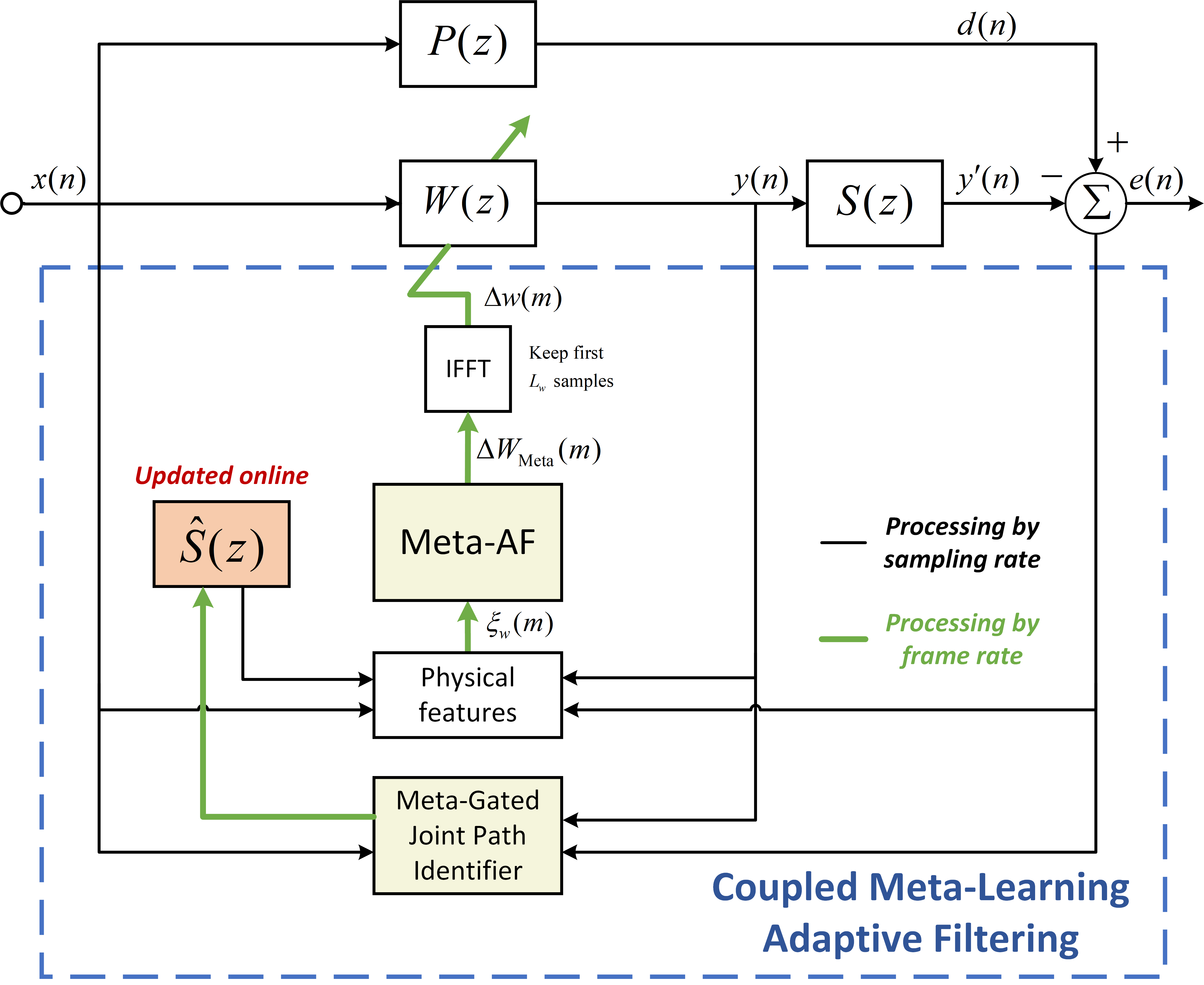}\vspace*{-0.2cm}
\caption{Proposed CoMeta-AF-ANC framework with a delayless dual-rate structure, where the control signal is generated sample by sample in the time domain while filter adaptation is performed at the frame rate. The control filter is updated by a Meta-AF optimizer, while MG-JPI performs online acoustic path tracking. The controller output provides the excitation for path identification, and the updated secondary path estimate is immediately used to reconstruct the Meta-AF controller features for subsequent control filter adaptation.}
\label{fig_3}
\end{figure}

\subsection{Delayless Meta-AF Control Filter Adaptation}
\label{sec:coupled_controller}

In CoMeta-AF-ANC, the control filter is implemented in the time domain to generate the control signal without block processing delay. At frame $m$, the coefficient vector $\mathbf w(m)=[w_1(m),w_2(m),\ldots,w_{L_w}(m)]^{\mathrm T}$ is held fixed over the corresponding hop interval, and the control signal is generated sample by sample as
\begin{equation}
    y(n)=\mathbf w^{\mathrm T}(m)\mathbf x(n),\qquad mH\le n<(m+1)H,
    \label{eq:delayless_control_signal}
\end{equation}

The control filter is adapted at the frame rate as
\begin{equation}
    \mathbf w(m+1)=\mathbf w(m)+\Delta\mathbf w(m).
    \label{eq:proposed_w_update}
\end{equation}

The frequency-domain control filter increment $\Delta W(m)$ is represented by a one-sided spectrum containing $N/2+1$ nonnegative-frequency bins. During the inverse transform, the corresponding negative-frequency components are implicitly reconstructed by Hermitian symmetry, yielding the real-valued time-domain increment as
\begin{equation}
    \Delta\mathbf w(m)=\operatorname{IFFT}\!\left\{\Delta W(m)\right\},
    \label{eq:w_td_increment}
\end{equation}
where only the first $L_w$ causal taps are retained after the inverse transform. 

Following the original Meta-AF formulation~\cite{casebeer2022meta}, the optimizer operates in the frequency domain, where adaptive filtering can exploit frequency-dependent signal statistics and perform per-bin adaptation efficiently. The frequency-domain increment is predicted by the Meta-AF optimizer as
\begin{equation}
    \Delta W(m)=g_{\phi_W}\!\left(\boldsymbol\xi_w(m)\right),
    \label{eq:w_optimizer_update}
\end{equation}
where $g_{\phi_W}(\cdot)$ denotes the Meta-AF controller optimizer parameterized by $\phi_W$.

Consistent with Eq.~\eqref{eq:metaaf_anc_features}, the input feature vector is defined as
\begin{equation}
    \boldsymbol\xi_w(m)=\left[-\left[X'^{+}(m)\right]^*E(m),\;X'^{+}(m),\;E(m)+\hat Y'^{+}(m),\;\hat Y'^{+}(m),\;E(m)\right]^{\mathrm T},
    \label{eq:w_optimizer_features}
\end{equation}
where the superscript $(\cdot)^+$ denotes the secondary-path-dependent features constructed using the latest online secondary path estimate.

\begin{figure}[!t]
\centering
\includegraphics[width=\columnwidth]{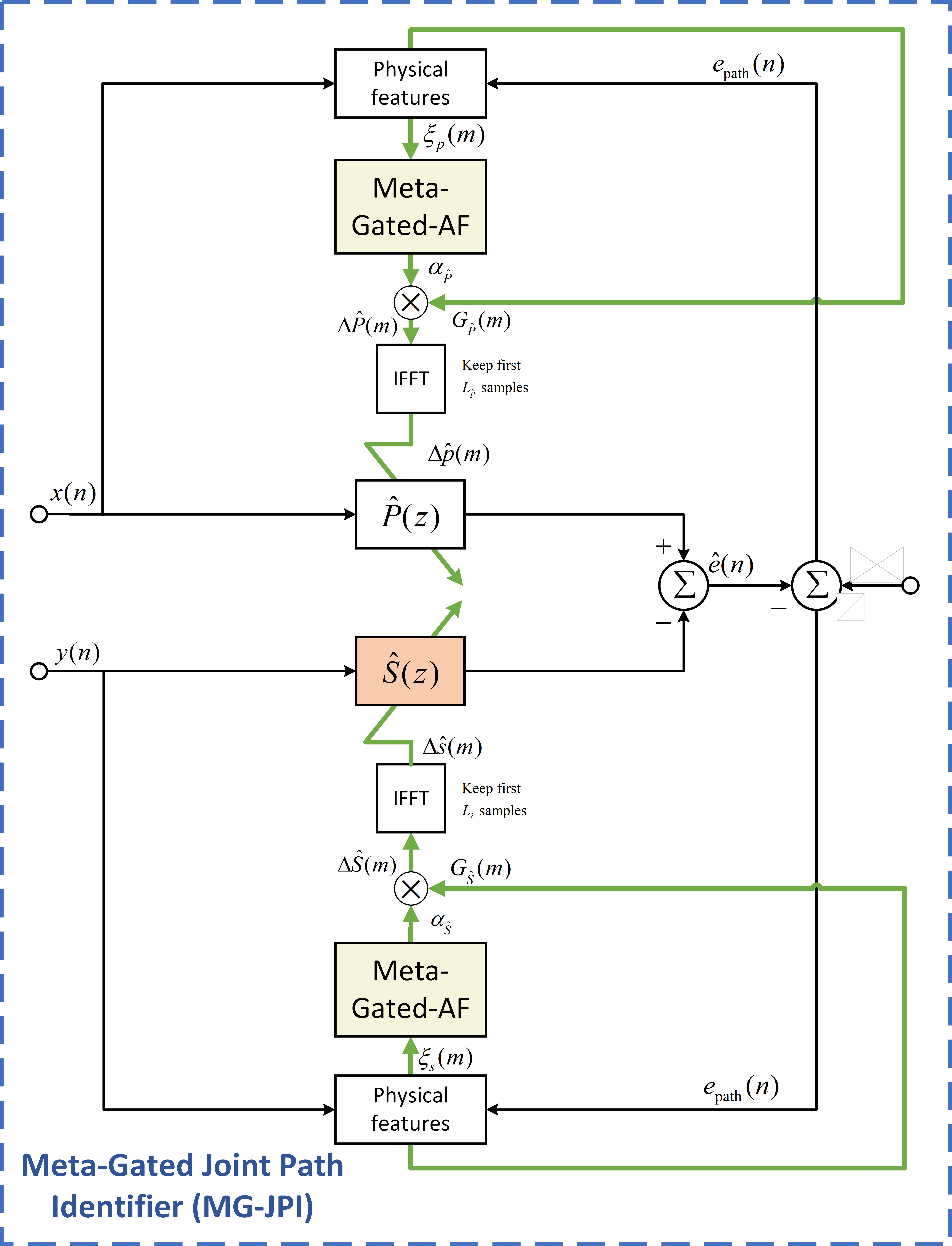}\vspace*{-0.3cm}
\caption{Block diagram of the proposed Meta-Gated Joint Path Identifier (MG-JPI). The primary and secondary paths are jointly modeled from a common residual, while the available reference, control, and error signals are used to construct the inputs to the corresponding Meta-Gated-AF optimizers. The optimizers learn frequency-dependent gates that modulate normalized path update directions, and the path models are updated in the time domain at the frame rate.}
\label{fig_4}
\end{figure}

\subsection{Meta-Gated Joint Path Identifier}
\label{sec:mgjpi_updates}
To maintain accurate secondary-path-dependent controller features under acoustic path variations, we propose MG-JPI, which jointly tracks the primary and secondary paths online using the available reference, control, and error signals without auxiliary noise. Its detailed signal flow is shown in Fig.~\ref{fig_4}. To jointly identify the two acoustic paths, MG-JPI constructs a common modeling residual. Let $\hat p(m)$ and $\hat s(m)$ denote the corresponding path estimates at frame $m$, and let $\hat e(n)$ denote the error predicted by these estimates. The joint modeling residual is expressed as
\begin{align}
    e_{\rm path}(n)
    &=e(n)-\hat e(n),\qquad mH\le n<(m+1)H \notag\\
    &=\big[p(n)*x(n)-s(n)*y(n)\big]
      -\big[\hat p(m)*x(n)-\hat s(m)*y(n)\big] \notag\\
    &=\big[p(n)-\hat p(m)\big]*x(n)
      -\big[s(n)-\hat s(m)\big]*y(n).
    \label{eq:path_residual_td}
\end{align}
Thus, $e_{\rm path}(n)$ contains the modeling errors of both paths under their respective excitations. Because the control signal is generated from the reference through the adaptive controller, the two excitation signals are generally correlated. Nevertheless, temporal variation of the control filter provides additional information for separating the primary and secondary path contributions in joint online modeling~\cite{hu2019online}. In MG-JPI, this variation is naturally provided by the continuously adapting controller, supporting joint path identification from the available ANC signals.

The corresponding path vectors $\hat{\mathbf p}(m)$ and $\hat{\mathbf s}(m)$ have lengths $L_{\hat p}$ and $L_{\hat s}$, respectively. The two path estimates are updated at the frame rate as
\begin{align}
    \hat{\mathbf p}(m+1)&=\hat{\mathbf p}(m)+\Delta\hat{\mathbf p}(m),\\
    \hat{\mathbf s}(m+1)&=\hat{\mathbf s}(m)+\Delta\hat{\mathbf s}(m).
    \label{eq:path_updates}
\end{align}

Let $\Delta\hat P(m)$ and $\Delta\hat S(m)$ denote the corresponding one-sided frequency-domain increments. Their real-valued time-domain increments are obtained using the same Hermitian-symmetric inverse transform as in Section~\ref{sec:coupled_controller}:
\begin{align}
    \Delta\hat{\mathbf p}(m)&=\operatorname{IFFT}\!\left\{\Delta\hat P(m)\right\},\\
    \Delta\hat{\mathbf s}(m)&=\operatorname{IFFT}\!\left\{\Delta\hat S(m)\right\}.
    \label{eq:path_td_increments}
\end{align}
After the inverse transforms, only the first $L_{\hat p}$ and $L_{\hat s}$ causal taps are retained for the primary and secondary path updates, respectively.

To determine the frequency-domain path increments, MG-JPI adopts a meta-gated adaptive filtering (Meta-Gated-AF) formulation that combines normalized update directions with learned frequency-dependent gating. Accordingly,
\begin{align}
    \Delta\hat P(m)&=\alpha_{\hat p}(m)G_{\hat p}(m),\\
    \Delta\hat S(m)&=\alpha_{\hat s}(m)G_{\hat s}(m),
    \label{eq:path_frequency_updates}
\end{align}
where the normalized update directions are
\begin{align}
    G_{\hat p}(m)&=\frac{X^*(m)E_{\rm path}(m)}{|X(m)|^2+\epsilon},
    \label{eq:path_update_direction_p}\\
    G_{\hat s}(m)&=-\frac{Y^*(m)E_{\rm path}(m)}{|Y(m)|^2+\epsilon},
    \label{eq:path_update_direction_s}
\end{align}
and $\epsilon>0$ is a small regularization constant. The corresponding frequency-dependent gates are generated by the two Meta-Gated-AF optimizers as
\begin{align}
    \alpha_{\hat p}(m)&=\sigma\!\left(\operatorname{Re}\!\left\{g_{\phi_P}\!\left(\boldsymbol\xi_{\hat p}(m)\right)\right\}\right),
    \label{eq:path_optimizer_output_p}\\
    \alpha_{\hat s}(m)&=\sigma\!\left(\operatorname{Re}\!\left\{g_{\phi_S}\!\left(\boldsymbol\xi_{\hat s}(m)\right)\right\}\right),
    \label{eq:path_optimizer_output_s}
\end{align}
where $g_{\phi_P}(\cdot)$ and $g_{\phi_S}(\cdot)$ denote the primary and secondary path Meta-Gated-AF optimizers parameterized by $\phi_P$ and $\phi_S$, respectively. The real part of each complex optimizer output is extracted before applying the sigmoid function $\sigma(\cdot)$ to obtain a real-valued gate. The sigmoid constraint keeps the gates positive and bounded, preventing reversal of the complex update direction and excessive amplification. Overall, the normalization reduces the sensitivity of the update magnitude to the instantaneous excitation power, while the learned gates provide frequency-dependent scaling of the normalized update directions.

Following the input design of Eq.~\eqref{eq:metaaf_anc_features}, the feature vectors for the primary and secondary path optimizers are
\begin{align}
    \boldsymbol\xi_{\hat p}(m)&=\left[-X^*(m)E_{\rm path}(m),\;X(m),\;E(m)+\hat S(m)Y(m),\;\hat P(m)X(m),\;E_{\rm path}(m)\right]^{\mathrm T},
    \label{eq:path_optimizer_feature_p}\\
    \boldsymbol\xi_{\hat s}(m)&=\left[Y^*(m)E_{\rm path}(m),\;Y(m),\;\hat P(m)X(m)-E(m),\;\hat S(m)Y(m),\;-E_{\rm path}(m)\right]^{\mathrm T},
    \label{eq:path_optimizer_feature_s}
\end{align}
where $\hat P(m)$, $\hat S(m)$, and $E_{\rm path}(m)$ denote the frequency-domain estimated primary path, estimated secondary path, and joint modeling residual, respectively.

\subsection{Coupled Online Recursion}
\label{sec:online_recursion}
The coupling is realized through sequential path and controller adaptation within each frame. The controller output generated during the current frame serves as an excitation for MG-JPI. After the path estimates are updated, the latest secondary path estimate is used to reconstruct the controller features before computing the next control filter update. The frame-$m$ error spectrum is used only for updates applied from frame $m+1$, thereby preserving causality. The complete online procedure of CoMeta-AF-ANC is summarized in Table~\ref{tab:cometaaf_pseudocode}.

\begin{table}[!t]
\centering
\small
\caption{Pseudo-code for the online CoMeta-AF-ANC recursion.}\vspace*{-0.3cm}
\begin{tabularx}{\linewidth}{@{}X@{}}
\toprule
\textbf{Initialization:} Initialize the control filter, primary path estimate, and all recurrent states to zero, and the secondary path estimate to a fixed nominal model.\\
\textbf{Input:} Reference signal $x(n)$, control signal $y(n)$, and error signal $e(n)$.\\
\midrule
\textbf{for} each frame $m$ \textbf{do}\\

\textbf{\# Delayless noise control (sampling rate):}\\
\hspace*{1em}\textbf{for} $n=mH,\ldots,(m+1)H-1$ \textbf{do}\\
\hspace*{2em}$y(n)\leftarrow\mathbf w^{\mathrm T}(m)\mathbf x(n)$ \hfill$\triangleright$ Generate the control signal.\\
\hspace*{2em}$e(n)\leftarrow d(n)-y(n)*s(n)$ \hfill$\triangleright$ Acquire the residual at the error microphone.\\
\hspace*{1em}\textbf{end for}\\

\textbf{\# Meta-Gated-AF joint path identification (frame rate):}\\
\hspace*{1em}Construct $\boldsymbol\xi_{\hat p}(m)$ and $\boldsymbol\xi_{\hat s}(m)$ using Eqs.~\eqref{eq:path_optimizer_feature_p} and~\eqref{eq:path_optimizer_feature_s}.\\
\hspace*{1em}$\alpha_{\hat p}(m)\leftarrow g_{\phi_P}(\boldsymbol\xi_{\hat p}(m))$ \hfill$\triangleright$ Predict the primary path gate.\\
\hspace*{1em}$\alpha_{\hat s}(m)\leftarrow g_{\phi_S}(\boldsymbol\xi_{\hat s}(m))$ \hfill$\triangleright$ Predict the secondary path gate.\\
\hspace*{1em}Compute $G_{\hat p}(m)$ and $G_{\hat s}(m)$ using Eqs.~\eqref{eq:path_update_direction_p} and~\eqref{eq:path_update_direction_s}.\\
\hspace*{1em}$\Delta\hat P(m)\leftarrow\alpha_{\hat p}(m)G_{\hat p}(m)$ \hfill$\triangleright$ Predict the primary path increment.\\
\hspace*{1em}$\Delta\hat S(m)\leftarrow\alpha_{\hat s}(m)G_{\hat s}(m)$ \hfill$\triangleright$ Predict the secondary path increment.\\
\hspace*{1em}$\Delta\hat{\mathbf p}(m)\leftarrow\operatorname{IFFT}\{\Delta\hat P(m)\}$ \hfill$\triangleright$ Convert to a time-domain update.\\
\hspace*{1em}$\Delta\hat{\mathbf s}(m)\leftarrow\operatorname{IFFT}\{\Delta\hat S(m)\}$ \hfill$\triangleright$ Convert to a time-domain update.\\
\hspace*{1em}$\hat{\mathbf p}(m+1)\leftarrow\hat{\mathbf p}(m)+\Delta\hat{\mathbf p}(m)$ \hfill$\triangleright$ Update the primary path estimate.\\
\hspace*{1em}$\hat{\mathbf s}(m+1)\leftarrow\hat{\mathbf s}(m)+\Delta\hat{\mathbf s}(m)$ \hfill$\triangleright$ Update the secondary path estimate.\\

\textbf{\# Meta-AF control filter adaptation (frame rate):}\\
\hspace*{1em}Construct $\boldsymbol\xi_w(m)$ using Eq.~\eqref{eq:w_optimizer_features} with the latest secondary path estimate.\\
\hspace*{1em}$\Delta W(m)\leftarrow g_{\phi_W}(\boldsymbol\xi_w(m))$ \hfill$\triangleright$ Predict the control filter increment.\\
\hspace*{1em}$\Delta\mathbf w(m)\leftarrow\operatorname{IFFT}\{\Delta W(m)\}$ \hfill$\triangleright$ Convert to a time-domain update.\\
\hspace*{1em}$\mathbf w(m+1)\leftarrow\mathbf w(m)+\Delta\mathbf w(m)$ \hfill$\triangleright$ Update the control filter.\\
\textbf{end for}\\
\bottomrule
\end{tabularx}
\label{tab:cometaaf_pseudocode}
\vspace*{-0.4cm}
\end{table}

\subsection{Network Architecture}
\label{sec:optimizer_architecture}
The Meta-AF optimizer and the two Meta-Gated-AF optimizers adopt the same complex-valued recurrent architecture, composed of fully connected (FC) layers, gated recurrent unit (GRU) layers, and rectified linear unit (ReLU) activations, but use independent parameter sets. Within each optimizer, the network parameters are shared across frequency bins, while each bin maintains its own recurrent states. 

For each optimizer, the complex-valued input feature vector is first magnitude-compressed while preserving its phase:
\begin{equation}
    \mathcal C\!\left(\boldsymbol\xi_q(m)\right)
    =\ln\!\left(1+\left|\boldsymbol\xi_q(m)\right|\right)
    \mathrm e^{i\angle \boldsymbol\xi_q(m)},
    \label{eq:complex_compression}
\end{equation}
where $q\in\{W,P,S\}$ indexes the control filter, primary path, and secondary path optimizers, respectively, and $i=\sqrt{-1}$. 

The compressed input is first projected by an FC layer followed by ReLU activation,
\begin{equation}
    \mathbf z_q(m)=\operatorname{ReLU}\!\left[\operatorname{FC}_1\!\left(\mathcal C\!\left(\boldsymbol\xi_q(m)\right)\right)\right].
    \label{eq:meta_optimizer_input_projection}
\end{equation}

The projected features are then processed by two stacked GRU layers,
\begin{align}
    \mathbf h_{q,1}(m)
    &=\operatorname{GRU}_1\!\left(\mathbf z_q(m),\mathbf h_{q,1}(m-1)\right),\\
    \mathbf h_{q,2}(m)
    &=\operatorname{GRU}_2\!\left(\mathbf h_{q,1}(m),\mathbf h_{q,2}(m-1)\right),
    \label{eq:meta_optimizer_grus}
\end{align}
where $\mathbf h_{q,1}(m)$ and $\mathbf h_{q,2}(m)$ are the recurrent states. The GRUs retain optimizer memory across frames, allowing each update to exploit information accumulated during adaptation. 

The final FC layers map the recurrent representation to the optimizer-specific complex output as
\begin{equation}
    g_{\phi_q}\!\left(\boldsymbol\xi_q(m)\right)
    =\operatorname{FC}_3\!\left(
    \operatorname{ReLU}\!\left[
    \operatorname{FC}_2\!\left(\mathbf h_{q,2}(m)\right)
    \right]\right),
    \qquad q\in\{W,P,S\}.
    \label{eq:meta_optimizer_output}
\end{equation}
For the controller optimizer, the complex output is used directly as the control filter increment $\Delta W(m)$. For the primary and secondary path optimizers, the real part of the corresponding complex output is passed through the sigmoid transformations in Eqs.~\eqref{eq:path_optimizer_output_p} and~\eqref{eq:path_optimizer_output_s} to obtain the bounded gates $\alpha_{\hat p}(m)$ and $\alpha_{\hat s}(m)$, respectively.

\begin{table}[!t]
\centering
\small
\caption{Architecture and complexity of each meta-optimizer.}\vspace*{-0.3cm}
\begin{tabularx}{\linewidth}{@{}lcccc@{}}
\toprule
\textbf{Layer} & \textbf{Input} & \textbf{Output}
& \textbf{Complex Params.} & \textbf{Complex MACs} \\
\midrule
Complex $\operatorname{FC}_1$ + ReLU & $5$ & $32$ & $192$ & $160$ \\
Complex $\operatorname{GRU}_1$ & $32$ & $32$ & $6{,}240$ & $6{,}144$ \\
Complex $\operatorname{GRU}_2$ & $32$ & $32$ & $6{,}240$ & $6{,}144$ \\
Complex $\operatorname{FC}_2$ + ReLU & $32$ & $32$ & $1{,}056$ & $1{,}024$ \\
Complex $\operatorname{FC}_3$ & $32$ & $1$ & $33$ & $32$ \\
\midrule
Total & -- & -- & $13{,}761$ & $13{,}504$ \\
\bottomrule
\end{tabularx}
\label{tab:meta_optimizer_architecture}
\vspace*{-0.3cm}
\end{table}

The layer configuration, parameter count, and multiply--accumulate (MAC) complexity of each optimizer are summarized in Table~\ref{tab:meta_optimizer_architecture}, where the MAC count is reported per frequency bin. Since the network parameters are shared across frequency bins, the three optimizers contain $0.04$ million complex parameters in total. With the experimental settings of $N=1024$, $H=512$, and $f_s=8$~kHz, their combined computational cost is approximately $324.7$ million complex MACs/s.

To assess online computational feasibility, the three optimizers are jointly evaluated on a single CPU thread of an AMD Ryzen Threadripper PRO 5975WX. The median inference time is $4.04$~ms per frame, substantially shorter than the $64$~ms hop interval, corresponding to a real-time factor (RTF) of approximately $0.063$. In practical implementations, the learned optimizers can also be deployed on a co-processor~\cite{wang2026predictive}, offloading their computation from the real-time controller while preserving sample-rate control signal generation.

\subsection{Joint Closed-Loop Training}
\label{sec:joint_training}
The three meta-optimizers $g_{\phi_W}$, $g_{\phi_P}$, and $g_{\phi_S}$ are jointly trained through the coupled online recursion in Section~\ref{sec:online_recursion}. A multi-frame loss over $M$ consecutive frames is adopted so that each learned update is evaluated based on its effect on subsequent closed-loop states and error signals rather than on one-step error reduction alone. The noise control loss is defined by
\begin{equation}
    \mathcal L_e=
    \frac{1}{M}\sum_{m=0}^{M-1}
    \ln\!\left(\|\mathbf e(m)\|^2\right),
    \label{eq:measured_error_loss}
\end{equation}
where $\mathbf e(m)$ collects the error samples in frame $m$, $\|\cdot\|$ denotes the Euclidean norm, and the logarithm compresses the dynamic range of the frame-level error energy during training. This loss enables unsupervised training directly from the measured error signal, eliminating the need for explicit labels and simplifying the data preparation. 

To ensure accurate secondary path tracking for constructing the control filter features, the secondary path modeling loss is defined by
\begin{equation}
    \mathcal L_S=
    \frac{1}{M}\sum_{m=0}^{M-1}
    \frac{\|\hat S(m)-S(m)\|^2}
    {\|S(m)\|^2},
    \label{eq:secondary_nmse_loss}
\end{equation}
where the norm is evaluated over the training frequency band. The true secondary path is used only to compute this loss during training and is not required during deployment. The overall training loss is
\begin{equation}
    \mathcal L=\mathcal L_e+\lambda_S\mathcal L_S,
    \label{eq:proposed_training_loss}
\end{equation}
where $\lambda_S$ balances noise control performance and secondary path modeling. No direct primary path modeling loss is imposed, since the primary path estimate supports joint residual separation and is optimized indirectly through the overall training objective.

\begin{table}[!t]
\centering
\small
\caption{Pseudo-code for the joint closed-loop training procedure.}\vspace*{-0.3cm}
\begin{tabularx}{\linewidth}{@{}X@{}}
\toprule
\textbf{Initialization:} Initialize the meta-optimizer parameters $\boldsymbol\phi=\{\phi_W,\phi_P,\phi_S\}$; initialize the control filter, primary path estimate, recurrent states, and signal histories to zero, and the secondary path estimate to a fixed nominal model.\\
\textbf{Input:} Sampled noise signals and time-varying acoustic paths.\\
\midrule

\textbf{for} each training segment \textbf{do}\\

\textbf{\# Closed-loop recursion:}\\
\hspace*{1em}Execute the online CoMeta-AF-ANC recursion in Table~\ref{tab:cometaaf_pseudocode} for $M$ consecutive frames.\\

\textbf{\# Multi-frame loss computation:}\\
\hspace*{1em}Compute $\mathcal L_e$ and $\mathcal L_S$ using Eqs.~\eqref{eq:measured_error_loss} and~\eqref{eq:secondary_nmse_loss}.\\
\hspace*{1em}$\mathcal L\leftarrow\mathcal L_e+\lambda_S\mathcal L_S$.
\hfill$\triangleright$ Form the training loss.\\

\textbf{\# Meta-optimizer update:}\\
\hspace*{1em}Backpropagate $\mathcal L$ over the $M$ consecutive frames and update $\phi_W$, $\phi_P$, and $\phi_S$.\\
\hspace*{1em}Detach the computation graph while retaining the adaptive and recurrent states.\\
\textbf{end for}\\

\bottomrule
\end{tabularx}
\label{tab:joint_training_pseudocode}
\vspace*{-0.4cm}
\end{table}

The joint training procedure is summarized in Table~\ref{tab:joint_training_pseudocode}. For each training segment, the coupled online recursion is executed for $M$ consecutive frames, after which the multi-frame loss is backpropagated through the recursion to jointly update $\phi_W$, $\phi_P$, and $\phi_S$. The computation graph is then detached, while the control filter, path estimates, recurrent states, and signal histories are carried forward to the next segment, implementing truncated backpropagation through time without resetting the online states. During online deployment, the optimizer parameters remain fixed while the adaptive and recurrent states continue to evolve online.

%% file: sections/04_experiments.tex
\section{Experimental Evaluation}
\label{sec:experiments}
In this section, a series of experiments is conducted to evaluate the efficacy of the proposed CoMeta-AF-ANC method on measured acoustic paths through comparisons with representative ANC methods. The evaluation first examines its tracking performance under time-varying acoustic paths, then assesses its generalization to unseen real-world noises and compares it with an initialization-based meta-learning method, and finally conducts an ablation study of MG-JPI and evaluates the acoustic path modeling performance.

\begin{figure*}[!t]
\centering
\includegraphics[width=5.5in]{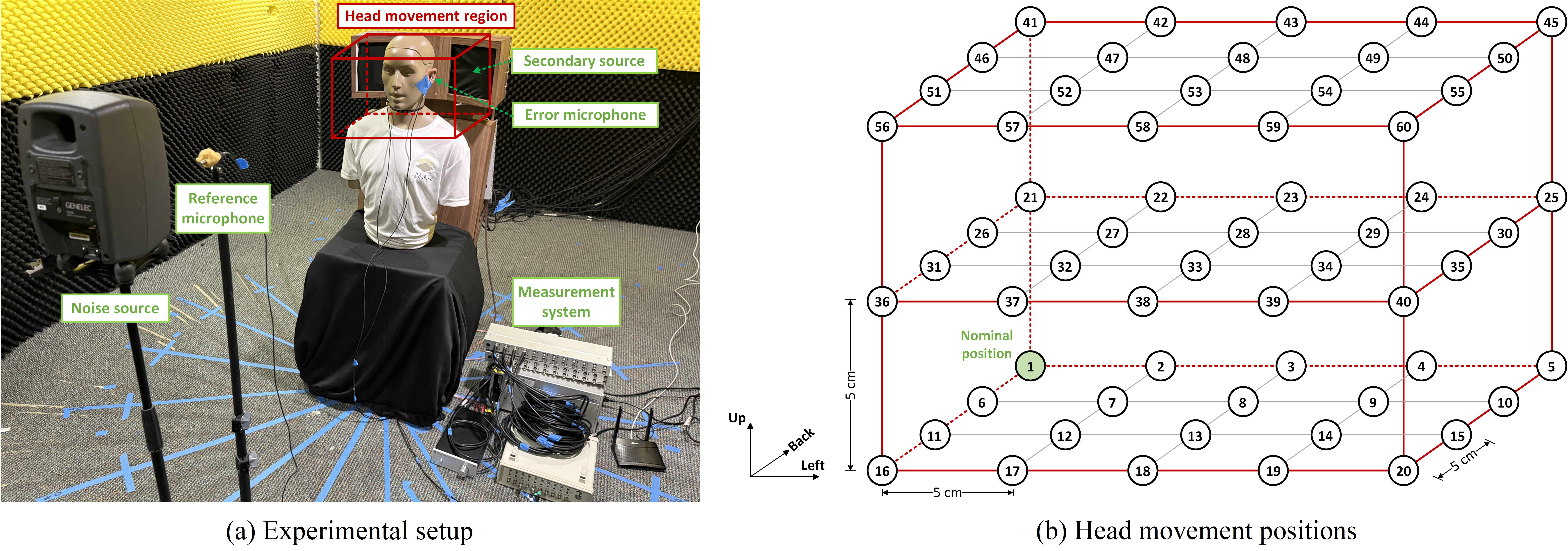}\vspace*{-0.3cm}
\caption{(a) Experimental setup of the headrest ANC system. (b) Measured head position grid within the head movement region, where adjacent positions are separated by $5$~cm along each spatial dimension.}
\label{fig:headrest_layout}
\vspace*{-0.3cm}
\end{figure*}

\subsection{Experimental Setup}
\label{sec:experimental_setup}
The experiments were conducted using a headrest ANC setup, in which the listener's head movements cause variations in both the primary and secondary acoustic paths. In this work, the left-side headrest ANC channel was used, consisting of one reference microphone, one secondary loudspeaker, and one error microphone, as illustrated in Fig.~\ref{fig:headrest_layout}(a). To characterize the acoustic path variations caused by head movement, the primary and secondary paths were measured at $60$ positions within the head movement region. Each position therefore corresponds to a measured primary--secondary path pair, denoted by $(P_j,S_j)$, where the subscript $j$ denotes the position index. These positions form a \(5\times4\times3\) three-dimensional grid with a uniform spacing of $5$~cm along the forward/backward, left/right, and up/down directions, as shown in Fig.~\ref{fig:headrest_layout}(b). Position $1$ is selected as the nominal position and serves as the initial state before any simulated head movement. 

The online processing settings used throughout the experiments are summarized in Table~\ref{tab:experimental_parameters}.

\begin{table}[!t]
\centering
\caption{Parameter settings used in the experimental evaluation.}\vspace*{-0.3cm}
\footnotesize
\renewcommand{\arraystretch}{1.03}
\begin{tabularx}{\linewidth}{@{}lXr@{}}
\toprule
\textbf{Symbol / Method} & \textbf{Definition} & \textbf{Value} \\
\midrule
\multicolumn{3}{@{}l}{\textbf{Online processing}} \\
$f_s$ & Sampling rate & $8$ kHz \\
$N$ & FFT/IFFT size & $1024$ \\
$H$ & Hop size & $512$ \\
$L_w$ & Control filter length & $512$ \\
$L_{\hat p}$ & Primary path estimate length & $512$ \\
$L_{\hat s}$ & Secondary path estimate length & $256$ \\
\midrule
\multicolumn{3}{@{}l}{\textbf{Dataset construction}} \\
$T_{\rm tr}$ & Path transition duration & $\{0,0.25,1,2,5,10\}$ s \\
-- & Position split (train/validation/test) & $44/6/10$ \\
-- & Instance duration & $30$ s  \\
-- & Background noise SNR  & $U(5,35)$ dB \\
\midrule
\multicolumn{3}{@{}l}{\textbf{Network training}} \\
$\lambda_S$ & Secondary path modeling loss weight & $1$ \\
$M$ & Backpropagation segment length & $28$ frames \\
-- & Target frequency band & $100$--$2000$ Hz \\
-- & Learning rate & $5\times10^{-5}$ \\
-- & Batch size & $10$ \\
-- & Global gradient clipping & $5$ \\
-- & Training iterations & $12000$ \\
\midrule
\multicolumn{3}{@{}l}{\textbf{Baseline parameters}} \\
\multicolumn{3}{@{}l}{\emph{FxLMS}} \\
& Control filter step size & $3\times10^{-6}$ \\
\addlinespace[1pt]
\multicolumn{3}{@{}l}{\emph{Delayless FD-FxNLMS}} \\
& Normalized control filter step size & $0.1$ \\
& Spectral power smoothing factor & $0.1$ \\
\addlinespace[1pt]
\multicolumn{3}{@{}l}{\emph{Zhang's OSPM / Co-initial MAML}} \\
& Control filter step size & $3\times10^{-5}$ \\
& Secondary path step size & $0.009$ \\
& Primary path step size & $5\times10^{-4}$ \\
& Auxiliary-noise power ratio & $0.09$ \\
\addlinespace[1pt]
\multicolumn{3}{@{}l}{\emph{Hu's OSPM}} \\
& Control filter step size & $8\times10^{-5}$ \\
& RLS forgetting factor & $0.99995$ \\
& RLS initialization constant & $10$ \\
\bottomrule
\end{tabularx}
\label{tab:experimental_parameters}
\vspace*{-0.3cm}
\end{table}

\subsubsection{Dataset Construction}
The training, validation, and test datasets are constructed by combining noise signals with time-varying acoustic paths derived from the measured headrest responses. The noise signals include both synthesized and real-world noises. The synthesized noises are generated as band-limited white noise with random bandwidths within the $100$--$2000$~Hz range, while the real-world noises are sourced from the \textit{UrbanSound8K} dataset~\cite{salamon2014dataset}. To improve robustness to background noise, additive noise is applied to both the reference and error signals during training. To construct the time-varying acoustic paths, each instance follows a three-position head movement scenario, starting from the nominal position and subsequently moving to two measured positions. Both sudden and gradual path transitions are simulated by varying the path transition duration \(T_{\rm tr}\), with gradual transitions generated using raised-cosine crossfading following~\cite{wefers2014efficient}. This dataset construction ensures that both the noise recordings and head positions used for testing are unseen during training. The detailed dataset settings are summarized in Table~\ref{tab:experimental_parameters}.

\subsubsection{Network Training}
The parameters of the three learned optimizers are jointly updated by a single Adam optimizer~\cite{kingma2014adam} following the closed-loop training procedure in Section~\ref{sec:joint_training}. Model selection is performed on held-out validation scenarios containing both synthesized and real-world noise signals. The network training settings are summarized in Table~\ref{tab:experimental_parameters}.

\subsubsection{Compared Methods}
The proposed CoMeta-AF-ANC method is compared with several representative ANC methods. The compared methods are summarized as follows:
\begin{enumerate}
\renewcommand{\labelenumi}{\theenumi)}
    \item \emph{Wiener Solution}: The Wiener solution provides an ideal reference for the attainable control performance when the acoustic paths are known~\cite{elliott2000signal}.
    \item \emph{FxLMS}: The conventional FxLMS controller uses a fixed nominal secondary path estimate throughout operation~\cite{yang2020stochastic}.
    \item \emph{Delayless FD-FxNLMS}: The control filter combines frequency-bin-normalized block updates with sample-by-sample time-domain filtering, using a fixed nominal secondary path estimate~\cite{elliott2000signal}.
    \item \emph{Zhang's OSPM}: An auxiliary-noise-based OSPM method injects an additional probing signal to support online secondary path identification~\cite{zhang2003robust}.
    \item \emph{Hu's OSPM}: An auxiliary-noise-free OSPM method identifies the secondary path from the available ANC signals without additional probing noise~\cite{hu2019online}.
    \item \emph{Co-initial MAML}: The control filter and secondary path estimate are jointly meta-initialized, providing an initialization-based meta-learning baseline~\cite{yang2026co}.
\end{enumerate}

For a fair comparison, the parameters of each baseline are selected by evaluating a predefined set of candidate values on the held-out validation set. Parameter settings that lead to unstable control are discarded, and the remaining setting with the highest average noise reduction is selected. The resulting parameter set is then fixed for all experiments and reported in Table~\ref{tab:experimental_parameters}.

\subsubsection{Evaluation Metrics}
The averaged noise reduction (NR) level, measured in dB, is defined as
\begin{equation}
    {\rm NR}(\mathcal I)=10\log_{10}\frac{\sum_{n\in\mathcal I}d^2(n)}
    {\sum_{n\in\mathcal I}e^2(n)},
    \label{eq:test_nr}
\end{equation}
where $\mathcal I$ denotes the selected evaluation interval. For the time-varying NR curves, frame-wise NR values are smoothed using a six-frame moving average.

The instability rate is defined as the percentage of unstable scenarios among all evaluated scenarios,
\begin{equation}
    R_{\rm inst}=\frac{N_{\rm inst}}{N_{\rm scen}}\times100\%,
    \label{eq:instability_rate}
\end{equation}
where $N_{\rm inst}$ and $N_{\rm scen}$ denote the numbers of unstable and total test scenarios, respectively. A scenario is considered unstable if non-finite values occur or if the steady-state NR over the final $2$~s after either path change falls below $0$~dB. 

\begin{figure}[!t]
\centering
\includegraphics[width=\columnwidth]{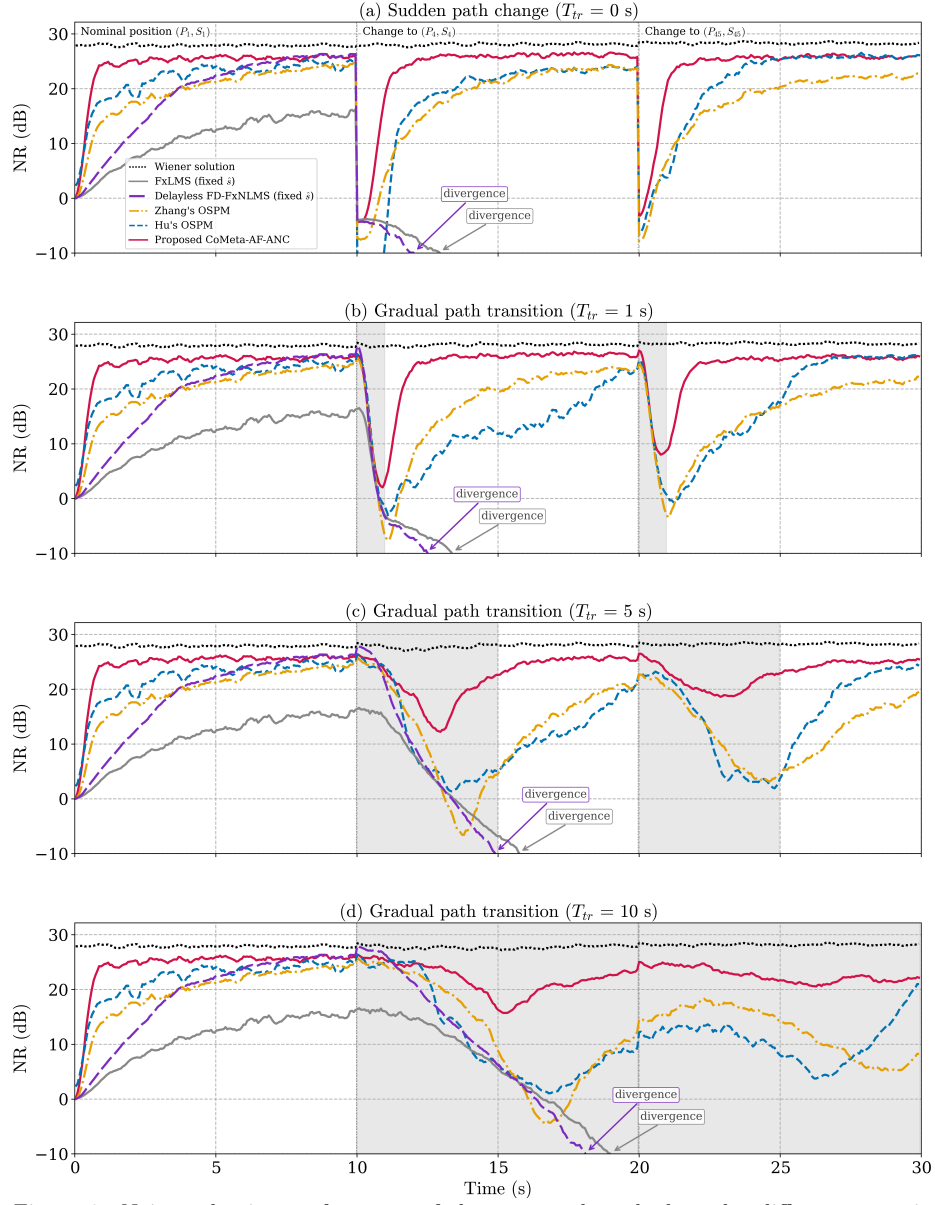}\vspace*{-0.3cm}
\caption{Noise reduction performance of the compared methods under different acoustic path transition durations for $500$--$1500$~Hz broadband noise in the head movement scenario $1\rightarrow4\rightarrow45$, with path transitions beginning at $10$ and $20$~s. The panels correspond to (a) a sudden transition and gradual transitions with $T_{\rm tr}=$ (b) $1$, (c) $5$, and (d) $10$~s. Gray shaded regions indicate the path transition intervals.}
\label{fig:dynamic_anc}
\vspace*{-0.3cm}
\end{figure}

\subsection{Performance Under Time-Varying Acoustic Paths}
\label{sec:results_dynamic_anc}
This subsection evaluates the performance of CoMeta-AF-ANC under time-varying acoustic paths with varying transition durations and unseen head movement scenarios. All experiments use $500$--$1500$~Hz broadband noise at an SNR of $30$~dB, with path transitions beginning at $10$ and $20$~s. 

The first experiment examines the effect of the path transition duration in the head movement scenario $1\rightarrow4\rightarrow45$, where position $1$ is the nominal position and positions $4$ and $45$ are held-out test positions. Fig.~\ref{fig:dynamic_anc} shows the noise control performance of the compared methods under different path transition durations, ranging from a sudden change to gradual transitions with durations of $1$, $5$, and $10$~s. For the sudden change case in Fig.~\ref{fig:dynamic_anc}(a), all methods converge under the nominal path, with CoMeta-AF-ANC approaching the Wiener reference more rapidly. After the path change, FxLMS and delayless FD-FxNLMS diverge due to secondary path mismatch, while Zhang's and Hu's OSPM methods undergo substantial NR degradation and recover relatively slowly. CoMeta-AF-ANC, in comparison, exhibits only a brief transient and restores more than $20$~dB of noise reduction within $1$~s after each change. Figs.~\ref{fig:dynamic_anc}(b)--(d) consider gradual transitions, which more closely represent continuous head movement. Across all transition durations, CoMeta-AF-ANC remains stable, maintains a higher NR, and effectively tracks the evolving acoustic paths. It also exhibits a smaller transient increase in residual noise during the transitions. These results demonstrate that CoMeta-AF-ANC provides both fast adaptation and robust tracking over a wide range of acoustic path transition durations.

\begin{table}[!t]
\centering
\caption{Performance for $90$ unseen head movement scenarios with sudden transitions for $500$--$1500$~Hz broadband noise. The instability rate is evaluated for all scenarios, while the NR during the first $3$~s following each transition is reported as the mean $\pm$ standard deviation across path change events within stable scenarios.}\vspace*{-0.3cm}
\small
\begin{tabular}{lcc}
\toprule
\textbf{Method}
& \textbf{Instability rate (\%) $\downarrow$}
& \textbf{Post-change NR ($0$--$3$~s) (dB) $\uparrow$} \\
\midrule
FxLMS & $53.3$ & $1.7\pm2.8$ \\
Delayless FD-FxNLMS & $65.6$ & $1.6\pm2.4$ \\
Zhang's OSPM & $6.7$ & $4.4\pm4.3$ \\
Hu's OSPM & $20.0$ & $0.5\pm8.6$ \\
\textbf{CoMeta-AF-ANC} & $\mathbf{0}$ & $\mathbf{7.9\pm3.5}$ \\
\bottomrule
\end{tabular}
\label{tab:multi_trajectory}
\vspace*{-0.3cm}
\end{table}

The second experiment evaluates the generalization of the proposed CoMeta-AF-ANC across different head movement scenarios. Each scenario starts from the nominal position and follows the sequence \(1\rightarrow a\rightarrow b\), with sudden path changes at each transition. Here, \(a\) and \(b\) are distinct positions selected from the $10$ held-out test positions, resulting in $90$ scenarios and $180$ path changes in total. As shown in Table~\ref{tab:multi_trajectory}, the instability rate in Eq.~\eqref{eq:instability_rate} is evaluated over all $90$ scenarios. CoMeta-AF-ANC remains stable for all scenarios, whereas FxLMS, delayless FD-FxNLMS, Zhang's OSPM, and Hu's OSPM exhibit instability rates of $53.3\%$, $65.6\%$, $6.7\%$, and $20.0\%$, respectively. For the stable scenarios, convergence after each path change is further assessed using the average NR over the first $3$~s. CoMeta-AF-ANC achieves the highest NR of $7.9\pm3.5$~dB, indicating faster convergence after path changes than the baseline methods. These results show that CoMeta-AF-ANC provides both improved stability and faster convergence across unseen head movement scenarios. Its consistent performance over diverse path changes further indicates reliable tracking even when the acoustic paths deviate substantially from the nominal condition.

\begin{figure}[!t]
\centering
\includegraphics[width=\columnwidth]{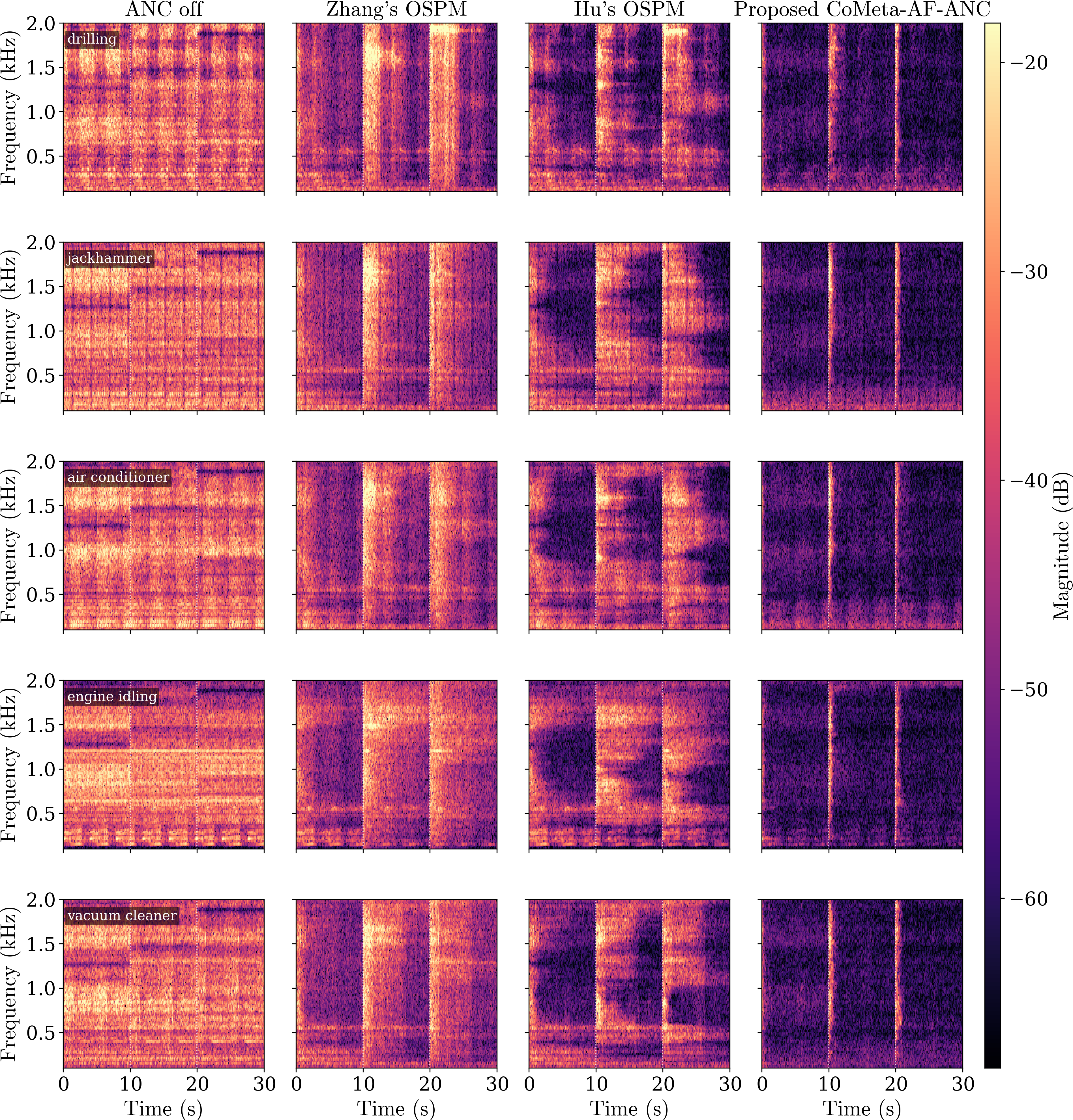}\vspace*{-0.3cm}
\caption{Spectrograms of the residual error signals for unseen drilling, jackhammer, air-conditioner, engine-idling, and vacuum-cleaner noises in the head movement scenario \(1\rightarrow4\rightarrow45\), with sudden path changes at $10$ and $20$~s.}
\label{fig:real_noise}
\vspace*{-0.3cm}
\end{figure}

\subsection{Real-World Noise Cancellation}
\label{sec:results_real_noise}
To evaluate the performance of the proposed CoMeta-AF-ANC under nonstationary real-world noise, five test recordings are selected, covering drilling, jackhammer, air-conditioner, engine-idling, and vacuum-cleaner noise, all of which are unseen during training. Fig.~\ref{fig:real_noise} presents the residual error spectrograms of the compared methods for each noise in the head movement scenario \(1\rightarrow4\rightarrow45\), with sudden path changes occurring at $10$ and $20$~s. As shown in Fig.~\ref{fig:real_noise}, CoMeta-AF-ANC maintains low residual energy across the target $100$--$2000$~Hz frequency band and rapidly restores effective attenuation after each path change. In contrast, Zhang's and Hu's OSPM methods show stronger and longer-lasting residual noise after each path change, with less uniform attenuation across frequency, indicating slower convergence and weaker tracking performance. The consistent results across all five unseen recordings, despite their diverse temporal and spectral characteristics, demonstrate that CoMeta-AF-ANC generalizes effectively to nonstationary real-world noise while maintaining fast tracking under acoustic path changes.

\begin{figure}[!t]
\centering
\includegraphics[width=\columnwidth]{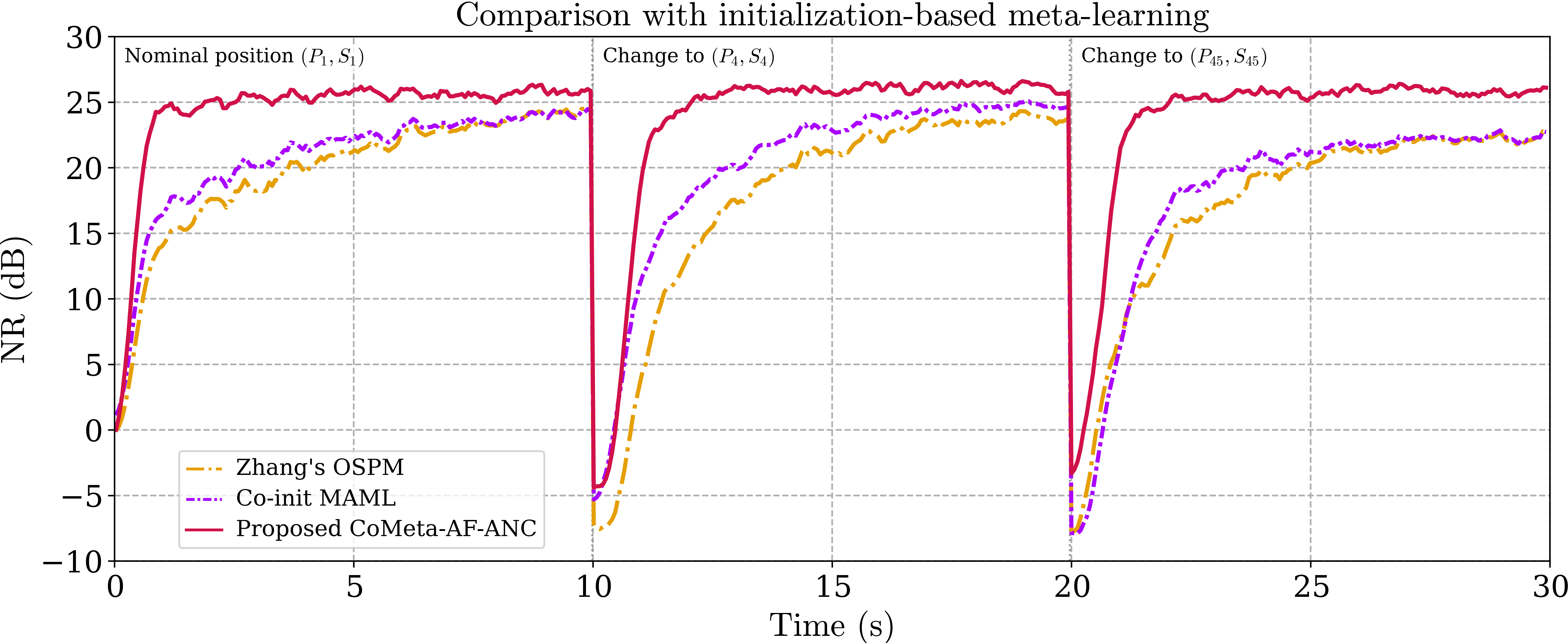}\vspace*{-0.3cm}
\caption{Comparison of noise reduction performance among CoMeta-AF-ANC, Co-initial MAML, and Zhang's OSPM for $500$--$1500$~Hz broadband noise in the head movement scenario $1\rightarrow4\rightarrow45$, with sudden path changes at $10$ and $20$~s.}
\label{fig:maml_comparison}
\vspace*{-0.3cm}
\end{figure}

\subsection{Comparison with Initialization-Based Meta-Learning}
\label{sec:results_maml}
This subsection compares the proposed CoMeta-AF-ANC with Co-initial MAML~\cite{yang2026co}, which meta-learns the initialization of both the control filter and the secondary path estimate, and subsequently uses the same online adaptation mechanism as Zhang's OSPM. To isolate the benefit of meta-learned initialization, Co-initial MAML and Zhang's OSPM use the same online adaptation parameters in Table~\ref{tab:experimental_parameters}, while Co-initial MAML is reinitialized to its meta-learned control filter and secondary path estimate after each path change. The experiment uses $500$--$1500$~Hz broadband noise and the head movement scenario \(1\rightarrow4\rightarrow45\), with sudden path changes at $10$ and $20$~s. As shown in Fig.~\ref{fig:maml_comparison}, Co-initial MAML consistently converges faster than Zhang's OSPM, both at startup and after each path change, while the two methods eventually approach similar NR levels. CoMeta-AF-ANC, however, achieves more than $20$~dB of noise reduction within approximately $1$~s after each change. Moreover, even after $10$~s of adaptation, Co-initial MAML remains approximately $2$--$4$~dB below CoMeta-AF-ANC. These results show that the initialization-based meta-learning method primarily provides a better starting point for the conventional adaptive algorithm, which accelerates convergence after initialization. By contrast, CoMeta-AF-ANC learns the update mechanism itself and keeps it active throughout online operation, allowing the controller to respond more quickly to path variations and achieve a higher NR thereafter.

\subsection{Ablation Study}
\label{sec:results_ablation}
To investigate the contribution of each component in MG-JPI, an ablation study is conducted using three variants of CoMeta-AF-ANC: Meta-AF-ANC, Meta-AF-ANC + Hu's OSPM, and CoMeta-AF-ANC without primary path modeling. Meta-AF-ANC uses the proposed delayless Meta-AF controller without online path identification. Meta-AF-ANC + Hu's OSPM combines the same controller with the conventional Hu's OSPM method. CoMeta-AF-ANC without primary path modeling retains the learned secondary path adaptation of MG-JPI but removes the primary path branch. The experiment uses $500$--$1500$~Hz broadband noise in the head movement scenario \(1\rightarrow4\rightarrow45\), with sudden path changes at $10$ and $20$~s.

\begin{figure}[!t]
\centering
\includegraphics[width=\columnwidth]{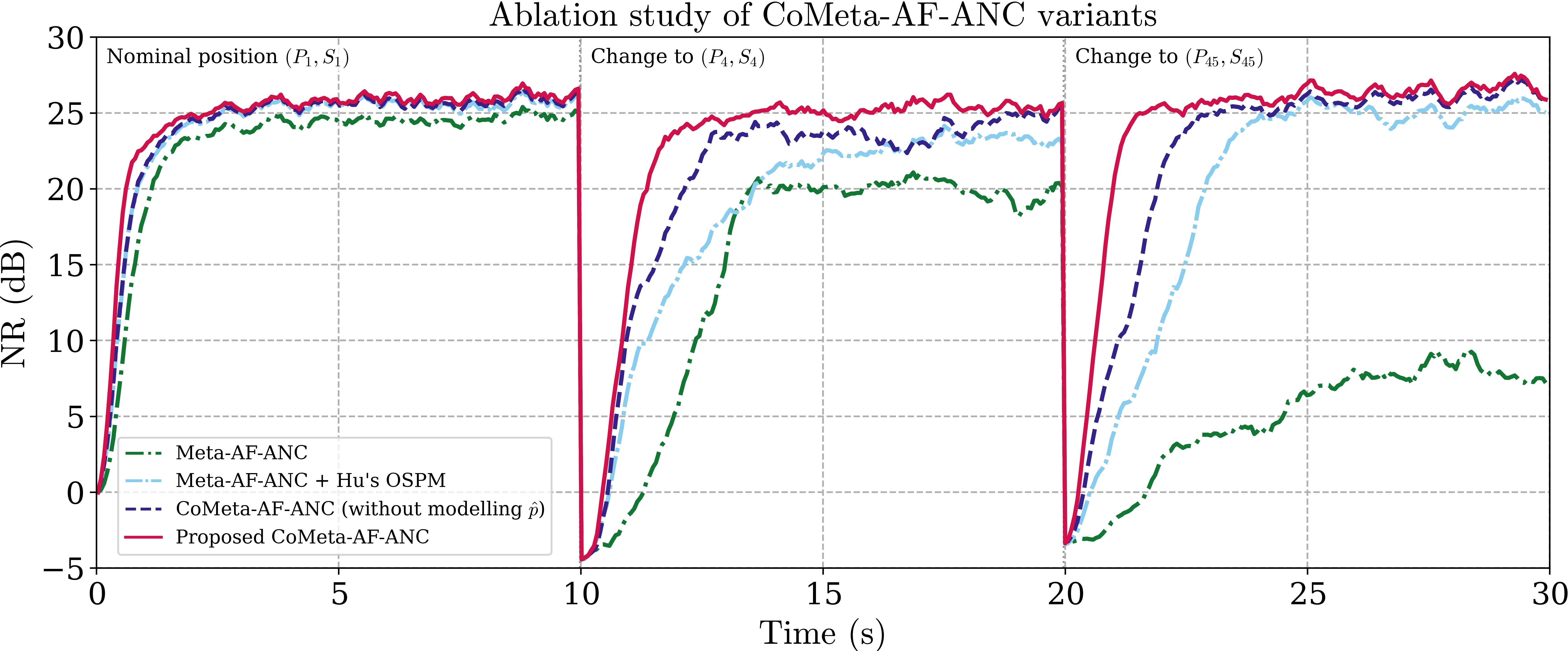}\vspace*{-0.3cm}
\caption{Comparison of CoMeta-AF-ANC variants for $500$--$1500$~Hz broadband noise in the head movement scenario \(1\rightarrow4\rightarrow45\), with sudden path changes at $10$ and $20$~s.}
\label{fig:ablation}
\vspace*{-0.3cm}
\end{figure}

As shown in Fig.~\ref{fig:ablation}, all four methods achieve comparable NR under the initially matched acoustic path, confirming that the Meta-AF controller provides effective noise control when the secondary path estimate is accurate. After the path changes, Meta-AF-ANC degrades substantially because the fixed secondary path estimate becomes mismatched. Adding Hu's OSPM improves robustness by enabling online secondary path modeling, but its tracking remains slower than that of CoMeta-AF-ANC without primary path modeling. This indicates that the learned secondary path adaptation in MG-JPI provides faster tracking than the conventional OSPM update. Moreover, CoMeta-AF-ANC without primary path modeling converges more slowly than the complete CoMeta-AF-ANC, showing that joint primary and secondary path modeling further improves the separation of their contributions in the residual signal and facilitates secondary path adaptation. Overall, the ablation results demonstrate that the performance gain of MG-JPI arises from both its learned path adaptation mechanism and the joint modeling strategy.

\begin{figure}[!t]
\centering
\includegraphics[width=\columnwidth]{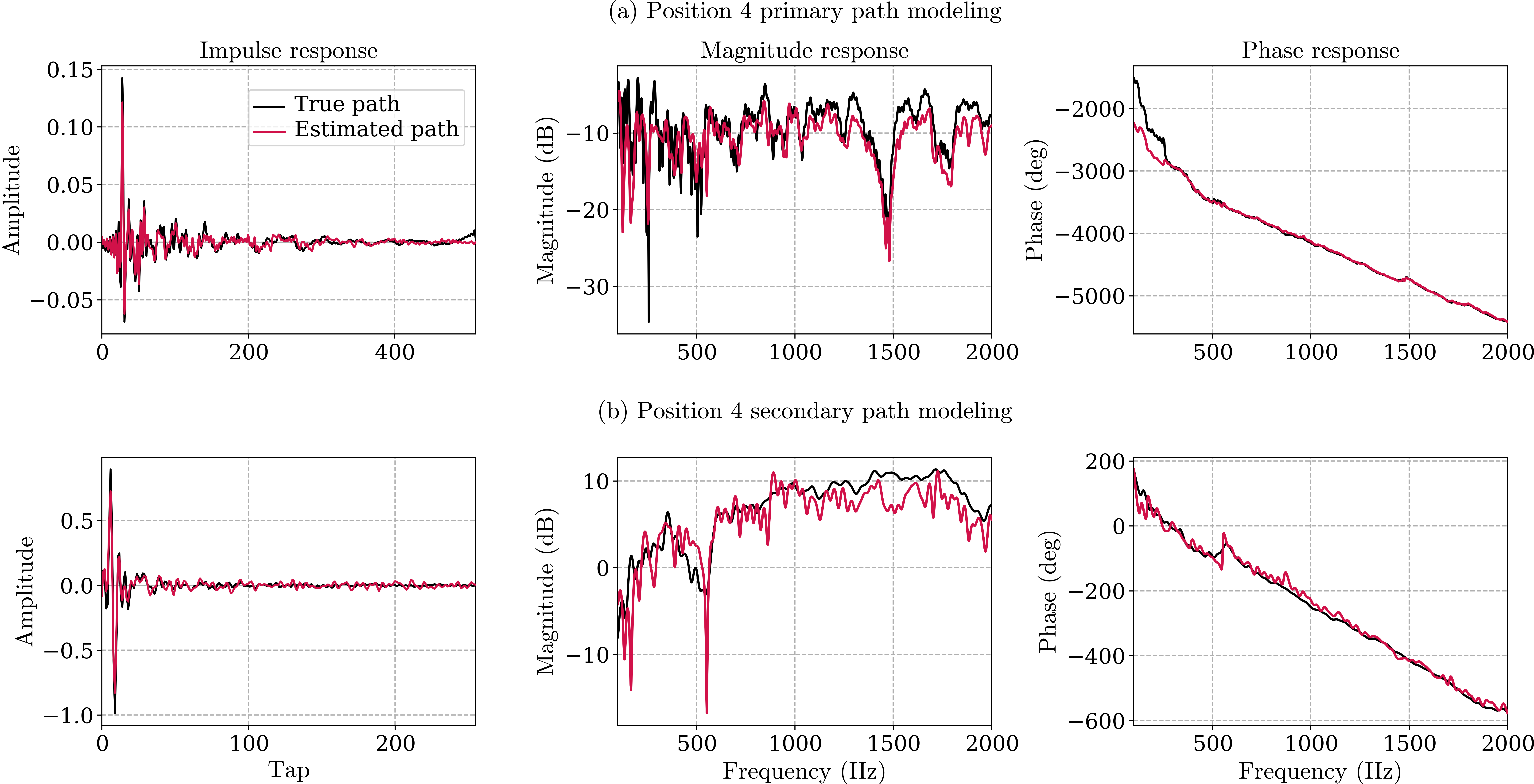}\vspace*{-0.3cm}
\caption{Acoustic path modeling results of CoMeta-AF-ANC at position $4$: (a) primary path and (b) secondary path, each shown in terms of impulse response, magnitude response, and phase response.}
\label{fig:path_modeling}
\vspace*{-0.3cm}
\end{figure}

\subsection{Acoustic Path Modeling Performance}
\label{sec:results_path_modeling}
To examine the acoustic modeling performance of CoMeta-AF-ANC, Fig.~\ref{fig:path_modeling} presents the true and estimated primary and secondary paths after adaptation to position $4$ from the nominal position $1$. The results are shown in both the time and frequency domains over the target $100$--$2000$~Hz control band. As shown in Fig.~\ref{fig:path_modeling}, the estimated primary and secondary paths reproduce the dominant impulse response structures and follow the corresponding true frequency responses well. The secondary path estimate therefore provides the Meta-AF controller with an accurate representation of the current secondary path characteristics. Although no direct primary path modeling loss is imposed, the primary path estimate also agrees well with the true response, indicating that this branch is learned indirectly through its role in separating the primary and secondary path contributions in the joint residual.

%% file: sections/05_conclusion.tex
\section{Conclusion}
\label{sec:conclusion}
In this paper, we propose CoMeta-AF-ANC, a coupled meta-adaptive filtering framework for ANC under time-varying acoustic paths. Unlike the existing Meta-AF-based ANC method that relies on a fixed secondary path model, CoMeta-AF-ANC couples control filter adaptation with online acoustic-path tracking within a joint meta-learning framework. The MG-JPI jointly tracks the primary and secondary paths without auxiliary noise, allowing the Meta-AF controller to continuously adapt to evolving acoustic paths. In addition, a delayless dual-rate realization performs learned adaptation at the frame rate while generating the control signal at the sampling rate in the time domain.

Experiments using measured headrest acoustic paths demonstrate that CoMeta-AF-ANC provides fast and robust adaptation under both sudden and gradual path variations, while maintaining stable operation across $90$ unseen head movement scenarios. Following sudden path changes, CoMeta-AF-ANC restores more than $20$~dB of noise reduction within approximately $1$~s. Comparative evaluations against representative ANC algorithms show that learning the online update mechanism provides a sustained adaptation advantage over initialization-based meta-learning, and that MG-JPI plays a critical role in maintaining effective control under secondary path variations. Moreover, CoMeta-AF-ANC generalizes well to nonstationary real-world noise unseen during training. Future work will investigate real-time implementation and evaluation under continuous listener movement.

%% file: refs.bib
@inproceedings{casebeer2021auto,
  title={Auto-DSP: Learning to optimize acoustic echo cancellers},
  author={Casebeer, Jonah and Bryan, Nicholas J and Smaragdis, Paris},
  booktitle={2021 IEEE Workshop on Applications of Signal Processing to Audio and Acoustics (WASPAA)},
  pages={291--295},
  year={2021},
  organization={IEEE}
}

@article{kuo1999active,
  title={Active noise control: a tutorial review},
  author={Kuo, Sen M and Morgan, Dennis R},
  journal={Proceedings of the IEEE},
  volume={87},
  number={6},
  pages={943--973},
  year={1999}
}

@article{shi2022selective,
  title={Selective fixed-filter active noise control based on convolutional neural network},
  author={Shi, Dongyuan and Lam, Bhan and Ooi, Kenneth and Shen, Xiaoyi and Gan, Woon-Seng},
  journal={Signal Processing},
  volume={190},
  pages={108317},
  year={2022},
  publisher={Elsevier}
}

@article{cheer2019application,
  title={The application of a multi-reference control strategy to noise cancelling headphones},
  author={Cheer, Jordan and Patel, Vinal and Fontana, Simone},
  journal={The Journal of the Acoustical Society of America},
  volume={145},
  number={5},
  pages={3095--3103},
  year={2019},
  publisher={AIP Publishing}
}

@article{luo2022hybrid,
  title={A hybrid sfanc-fxnlms algorithm for active noise control based on deep learning},
  author={Luo, Zhengding and Shi, Dongyuan and Gan, Woon-Seng},
  journal={IEEE Signal Processing Letters},
  volume={29},
  pages={1102--1106},
  year={2022},
  publisher={IEEE}
}

@inproceedings{salamon2014dataset,
  title={A dataset and taxonomy for urban sound research},
  author={Salamon, Justin and Jacoby, Christopher and Bello, Juan Pablo},
  booktitle={Proceedings of the 22nd ACM international conference on Multimedia},
  pages={1041--1044},
  year={2014}
}

@article{chang2022multi,
  title={Multi-functional active noise control system on headrest of airplane seat},
  author={Chang, Cheng-Yuan and Chuang, Chia-Tseng and Kuo, Sen M and Lin, Chia-Hao},
  journal={Mechanical Systems and Signal Processing},
  volume={167},
  pages={108552},
  year={2022},
  publisher={Elsevier}
}

@article{zhang2021deep,
  title={Deep ANC: A deep learning approach to active noise control},
  author={Zhang, Hao and Wang, DeLiang},
  journal={Neural Networks},
  volume={141},
  pages={1--10},
  year={2021},
  publisher={Elsevier}
}

@article{kajikawa2012recent,
  title={Recent advances on active noise control: open issues and innovative applications},
  author={Kajikawa, Yoshinobu and Gan, Woon-Seng and Kuo, Sen M},
  journal={APSIPA Transactions on Signal and Information Processing},
  volume={1},
  pages={e3},
  year={2012},
  publisher={Cambridge University Press}
}

@inproceedings{kingma2014adam,
  title={Adam: A Method for Stochastic Optimization},
  author={Kingma, Diederik P. and Ba, Jimmy},
  booktitle={International Conference on Learning Representations (ICLR)},
  year={2015}
}

@inproceedings{liebich2018signal,
  title={Signal processing challenges for active noise cancellation headphones},
  author={Liebich, Stefan and Fabry, Johannes and Jax, Peter and Vary, Peter},
  booktitle={Speech Communication; 13th ITG-Symposium},
  pages={1--5},
  year={2018},
  organization={VDE}
}

@book{elliott2000signal,
  title={Signal processing for active control},
  author={Elliott, Stephen},
  year={2000},
  publisher={Elsevier}
}

@article{aboutiman2025subjective,
  title={Subjective perception analysis of active noise control algorithms in an encapsulated structure: An experimental study},
  author={Aboutiman, Alkahf and Rachman, Zulfi and Oberman, Tin and Alletta, Francesco and Kang, Jian and Karimi, Hamid Reza and Ripamonti, Francesco},
  journal={Applied Acoustics},
  volume={239},
  pages={110823},
  year={2025},
  publisher={Elsevier}
}

@article{luo2025mssp,
  title={Deep learning-based Generative Fixed-Filter Active Noise Control: Transferability and implementation},
  author={Luo, Zhengding and Ji, Junwei and Wang, Boxiang and Shi, Dongyuan and Ma, Haozhe and Gan, Woon-Seng},
  journal={Mechanical Systems and Signal Processing},
  volume={238},
  pages={113207},
  year={2025},
  publisher={Elsevier}
}

@article{yang2020stochastic,
  title={Stochastic analysis of the filtered-x LMS algorithm for active noise control},
  author={Yang, Feiran and Guo, Jianfeng and Yang, Jun},
  journal={IEEE/ACM Transactions on Audio, Speech, and Language Processing},
  volume={28},
  pages={2252--2266},
  year={2020},
  publisher={IEEE}
}

@article{wang2026spatial,
  title={Spatial-Frequency Cued Generative Fixed-Filter Active Noise Control Based on Deep Learning in Reverberant Environments},
  author={Wang, Boxiang and Li, Haowen and Shi, Dongyuan and Ji, Junwei and Yang, Ziyi and Luo, Zhengding and Gan, Woon-Seng},
  journal={Signal Processing},
  pages={110682},
  year={2026},
  publisher={Elsevier}
}

@article{bai2026adaptive,
  title={An adaptive deep neural network for active road noise control},
  author={Bai, Lu and Xue, Jinpei and Lian, Siyuan and He, Yiming and Liang, Yayun and Rao, Li and Wang, Shuping and Lu, Jing},
  journal={The Journal of the Acoustical Society of America},
  volume={159},
  number={4},
  pages={3674--3685},
  year={2026},
  publisher={AIP Publishing}
}

@article{zhang2018active,
  title={Active noise control over space: A wave domain approach},
  author={Zhang, Jihui and Abhayapala, Thushara D and Zhang, Wen and Samarasinghe, Prasanga N and Jiang, Shouda},
  journal={IEEE/ACM Transactions on audio, speech, and language processing},
  volume={26},
  number={4},
  pages={774--786},
  year={2018},
  publisher={IEEE}
}

@inproceedings{park2023had,
  title={HAD-ANC: A Hybrid System Comprising an Adaptive Filter and Deep Neural Networks for Active Noise Control.},
  author={Park, JungPhil and Choi, Jeong-Hwan and Kim, Yungyeo and Chang, Joon-Hyuk},
  booktitle={INTERSPEECH},
  pages={2513--2517},
  year={2023}
}

@article{oh2026head,
  title={Head-tracking active road noise control by real-time path adaptation using neural networks},
  author={Oh, Jun Young and Kim, Ju Young and Oh, Chi Sung and Doe, Ji Hyun and Kang, Yeon June},
  journal={Mechanical Systems and Signal Processing},
  volume={250},
  pages={114106},
  year={2026},
  publisher={Elsevier}
}

@INPROCEEDINGS{yang2026co,
  author={Yang, Ziyi and Rao, Li and Luo, Zhengding and Shi, Dongyuan and Huang, Qirui and Gan, Woon-Seng},
  booktitle={ICASSP 2026 - 2026 IEEE International Conference on Acoustics, Speech and Signal Processing (ICASSP)}, 
  title={Co-Initialization of Control Filter and Secondary Path via Meta-Learning for Active Noise Control}, 
  year={2026},
  volume={},
  number={},
  pages={15297-15301}}

@article{cha2023dnoisenet,
  title={DNoiseNet: Deep learning-based feedback active noise control in various noisy environments},
  author={Cha, Young-Jin and Mostafavi, Alireza and Benipal, Sukhpreet S},
  journal={Engineering Applications of Artificial Intelligence},
  volume={121},
  pages={105971},
  year={2023},
  publisher={Elsevier}
}

@article{hu2019online,
  title={Online multi-channel secondary path modeling in active noise control without auxiliary noise},
  author={Hu, Meiling and Xue, Jinpei and Lu, Jing},
  journal={The Journal of the Acoustical Society of America},
  volume={146},
  number={4},
  pages={2590--2595},
  year={2019},
  publisher={AIP Publishing}
}

@inproceedings{lian2023frequency,
  title={Frequency domain online secondary path modelling for active noise control without auxiliary noise},
  author={Lian, Siyuan and Li, Tianyou and Wang, Shuping and Lu, Jing and Gu, Jincheng and Hu, Yuxiang and Zhu, Changbao},
  booktitle={INTER-NOISE and NOISE-CON Congress and Conference Proceedings},
  volume={268},
  number={5},
  pages={3041--3049},
  year={2023},
  organization={Institute of Noise Control Engineering}
}

@article{feng2025meta,
  title={Meta-learning-based delayless subband adaptive filter using complex self-attention for active noise control},
  author={Feng, Pengxing and So, Hing Cheung},
  journal={Neurocomputing},
  volume={650},
  pages={130637},
  year={2025},
  publisher={Elsevier}
}

@article{casebeer2022meta,
  title={Meta-AF: Meta-learning for adaptive filters},
  author={Casebeer, Jonah and Bryan, Nicholas J and Smaragdis, Paris},
  journal={IEEE/ACM Transactions on Audio, Speech, and Language Processing},
  volume={31},
  pages={355--370},
  year={2022},
  publisher={IEEE}
}

@article{li2026practical,
  title={A practical data-driven step-size selection method for adaptive active noise control based on modified meta-learning},
  author={Li, Luyuan and Su, Xiruo and Shi, Dongyuan and Chen, Jie and Gan, Woon-seng},
  journal={IEEE Signal Processing Letters},
  year={2026},
  publisher={IEEE}
}

@article{shi2021fast,
  title={Fast adaptive active noise control based on modified model-agnostic meta-learning algorithm},
  author={Shi, Dongyuan and Gan, Woon-Seng and Lam, Bhan and Ooi, Kenneth},
  journal={IEEE Signal Processing Letters},
  volume={28},
  pages={593--597},
  year={2021},
  publisher={IEEE}
}

@article{shen2025data,
  title={Data-driven method to accelerate convergence of adaptive hybrid active noise control: Two-stage model-agnostic meta-learning},
  author={Shen, Xiaoyi and Shi, Dongyuan and Gan, Woon-Seng},
  journal={IEEE Signal Processing Letters},
  year={2025},
  publisher={IEEE}
}

@article{eriksson1989use,
  title={Use of random noise for on-line transducer modeling in an adaptive active attenuation system},
  author={Eriksson, Larry J and Allie, Mark C},
  journal={The Journal of the Acoustical Society of America},
  volume={85},
  number={2},
  pages={797--802},
  year={1989},
  publisher={Acoustical Society of America}
}

@article{pradhan20205,
  title={A 5-stage active control method with online secondary path modelling using decorrelated control signal},
  author={Pradhan, Somanath and Qiu, Xiaojun},
  journal={Applied Acoustics},
  volume={164},
  pages={107252},
  year={2020},
  publisher={Elsevier}
}

@inproceedings{xiao2025soft,
  title={Soft-Constrained Spatially Selective Active Noise Control for Open-Fitting Hearables},
  author={Xiao, Tong and Roden, Reinhild and Blau, Matthias and Doclo, Simon},
  booktitle={2025 IEEE Workshop on Applications of Signal Processing to Audio and Acoustics (WASPAA)},
  pages={1--5},
  year={2025},
  organization={IEEE}
}

@article{iotov2025speech,
  title={Speech Prediction in ANC Headphones for Improved Attenuation: New Methods and Perceptual Study},
  author={Iotov, Yurii and Elofsson, Rasmus and N{\o}rholm, Sidsel Marie and McCutcheon, Peter John and Christensen, Mads Gr{\ae}sb{\o}ll},
  journal={IEEE Transactions on Audio, Speech and Language Processing},
  year={2025},
  publisher={IEEE}
}

@article{zhang2003robust,
  title={A robust online secondary path modeling method with auxiliary noise power scheduling strategy and norm constraint manipulation},
  author={Zhang, Ming and Lan, Hui and Ser, Wee},
  journal={IEEE Transactions on Speech and Audio Processing},
  volume={11},
  number={1},
  pages={45--53},
  year={2003},
  publisher={IEEE}
}

@inproceedings{zhang2024active,
  title={An active noise control system based on soundfield interpolation using a physics-informed neural network},
  author={Zhang, Yile Angela and Ma, Fei and Abhayapala, Thushara D and Samarasinghe, Prasanga N and Bastine, Amy},
  booktitle={ICASSP 2024-2024 IEEE International Conference on Acoustics, Speech and Signal Processing (ICASSP)},
  pages={506--510},
  year={2024},
  organization={IEEE}
}

@article{wang2026predictive,
  title={Predictive Directional Selective Fixed-Filter Active Noise Control for Moving Sources via a Convolutional Recurrent Neural Network},
  author={Wang, Boxiang and Luo, Zhengding and Shi, Dongyuan and Ji, Junwei and Su, Xiruo and Gan, Woon-Seng},
  journal={arXiv preprint arXiv:2604.23144},
  year={2026}
}

@inproceedings{wefers2014efficient,
  title={Efficient time-varying {FIR} filtering using crossfading implemented in the {DFT} domain},
  author={Wefers, Frank and Vorl{\"a}nder, Michael},
  booktitle={Proceedings of Forum Acusticum 2014},
  pages={1--6},
  address={Krakow, Poland},
  year={2014},
  organization={European Acoustics Association}
}

@article{tang2023stability,
  title={Stability guaranteed active noise control: Algorithms and applications},
  author={Tang, Yu and Zhang, Hongwei and Zhang, Yuting},
  journal={IEEE Transactions on Control Systems Technology},
  volume={31},
  number={4},
  pages={1720--1732},
  year={2023},
  publisher={IEEE}
}

@article{wen2026asynchronous,
  title={An asynchronous hierarchical dual-population framework for collaborative active noise control with online secondary-path modeling},
  author={Wen, Pengwei and Meng, Xu and Qu, Boyang and Yan, Li and Chai, Xuzhao and Zhao, Haiquan and Liang, Jing},
  journal={Expert Systems with Applications},
  volume={329},
  pages={132986},
  year={2026},
  publisher={Elsevier}
}

@article{ma2026robust,
  title={A robust feedforward active noise control system with simultaneous online secondary-and feedback-path modeling},
  author={Ma, Yaping and Xiao, Yegui and Wu, Wenyi and Ma, Liying and Khorasani, Khashayar},
  journal={Signal Processing},
  pages={110818},
  year={2026},
  publisher={Elsevier}
}

@article{wang2025online,
  title={Online secondary path modeling algorithm without auxiliary noise for narrowband active noise control},
  author={Wang, Cong and Wu, Ming and Guo, Jianfeng and Yang, Jun},
  journal={Signal Processing},
  volume={227},
  pages={109737},
  year={2025},
  publisher={Elsevier}
}
